\documentclass[twocolumn]{article}
\usepackage{amsmath}
\usepackage{amsfonts}

\usepackage{fullpage}

\begin{document}
\newcommand{\captionfontstyle}{\bf\footnotesize}
\newcommand{\Xng}[2]{{#1$\times$#2}}
\newcommand{\mytheoremcounter}{section}
\newcommand{\cheapHack}{ }

\newcommand{\myem}[1]{{\bf #1}}
\newcommand{\reccr}[1]{\overline{\mathrm{cr}}(#1)}
\newcommand{\regcr}[1]{\mathrm{cr}(#1)}
\newcommand{\NLclass}{\mathbf{NL}}
\newcommand{\NCclass}{\mathbf{NC}}
\newcommand{\coNLclass}{\mathbf{co\!-\!NL}}
\newcommand{\Lclass}{\mathbf{L}}
\newcommand{\Pclass}{\mathbf{P}}
\newcommand{\NPclass}{\mathbf{NP}}
\newcommand{\coNPclass}{\mathbf{co\!-\!NP}}
\newcommand{\PPclass}{\mathbf{PP}}
\newcommand{\BPPclass}{\mathbf{BPP}}
\newcommand{\ZPPclass}{\mathbf{ZPP}}
\newcommand{\RPclass}{\mathbf{RP}}
\newcommand{\coRPclass}{\mathbf{co\!-\!RP}}
\newcommand{\SigmaP}[1]{\mathbf{\Sigma_{#1}P}}
\newcommand{\PHclass}{\mathbf{PH}}
\newcommand{\PSPACE}{\mathbf{PSPACE}}
\newcommand{\RSPACE}{\mathbf{RSPACE}}
\newcommand{\DSPACE}{\mathbf{DSPACE}}
\newcommand{\imin}{{\mathit{min}}}
\newcommand{\imax}{{\mathit{max}}}
\newcommand{\uvec}{\vec{u}}
\newcommand{\vvec}{\vec{v}}
\newcommand{\xvec}{\vec{x}}
\newcommand{\Sigmap}{\Sigma^\prime}
\newcommand{\alphap}{\alpha^\prime}
\newcommand{\betap}{\beta^\prime}
\newcommand{\gammap}{\gamma^\prime}
\newcommand{\mup}{\mu^\prime}
\newcommand{\mupp}{\mu^{\prime\prime}}
\newcommand{\nubar}{{\overline{\nu}}}
\newcommand{\nubars}{\nubar^*}
\newcommand{\nus}{\nu^*}
\newcommand{\betas}{\beta^*}
\newcommand{\deltas}{\delta^*}
\newcommand{\deltap}{\delta^\prime}
\newcommand{\deltapp}{\delta^{\prime\prime}}
\newcommand{\lambdabar}{\overline{\lambda}}
\newcommand{\lambdap}{\lambda^\prime}
\newcommand{\lambdapp}{\lambda^{\prime\prime}}
\newcommand{\pip}{\pi^\prime}
\newcommand{\pipp}{{\pi^{\prime\prime}}}
\newcommand{\Psip}{\Psi^\prime}
\newcommand{\psip}{\psi^\prime}
\newcommand{\phip}{\phi^\prime}
\newcommand{\sigmap}{\sigma^\prime}
\newcommand{\etap}{\eta^\prime}
\newcommand{\etapp}{{\eta^{\prime\prime}}}
\newcommand{\epsilonp}{{\epsilon^\prime}}
\newcommand{\epsilonpp}{\epsilon^{\prime\prime}}
\newcommand{\epsilonppp}{\epsilon^{\prime\prime\prime}}
\newcommand{\ap}{a^\prime}
\newcommand{\bp}{b^\prime}
\newcommand{\cp}{c^\prime}
\newcommand{\up}{u^\prime}
\newcommand{\fp}{f^\prime}
\newcommand{\ep}{e^\prime}
\newcommand{\gp}{g^\prime}
\newcommand{\mpr}{m^\prime}
\newcommand{\np}{n^\prime}
\newcommand{\op}{o^\prime}
\newcommand{\qp}{q^\prime}
\newcommand{\rp}{r^\prime}
\newcommand{\gpp}{g^{\prime\prime}}
\newcommand{\epp}{e^{\prime\prime}}
\newcommand{\qpp}{q^{\prime\prime}}
\newcommand{\rpp}{r^{\prime\prime}}
\newcommand{\vp}{v^\prime}
\newcommand{\ypp}{y^{\prime\prime}}
\newcommand{\xpp}{x^{\prime\prime}}
\newcommand{\zp}{z^\prime}
\newcommand{\hp}{{h^\prime}}
\newcommand{\lp}{{l^\prime}}
\newcommand{\zpp}{{z^{\prime\prime}}}
\newcommand{\kp}{{k^\prime}}
\newcommand{\Dp}{{D^\prime}}
\newcommand{\Pp}{P^\prime}
\newcommand{\Ppp}{P^{\prime\prime}}
\newcommand{\Qp}{Q^\prime}
\newcommand{\Qpp}{Q^{\prime\prime}}
\newcommand{\Tp}{T^\prime}
\newcommand{\Ep}{E^\prime}
\newcommand{\Dbar}{\overline{D}}
\newcommand{\Lp}{L^\prime}
\newcommand{\Lpp}{L^{\prime\prime}}
\newcommand{\Lbar}{\overline{L}}
\newcommand{\Mp}{M^\prime}
\newcommand{\Mpp}{M^{\prime\prime}}
\newcommand{\Mbar}{\overline{M}}
\newcommand{\Np}{N^\prime}
\newcommand{\Npp}{N^{\prime\prime}}
\newcommand{\Rp}{R^\prime}
\newcommand{\xp}{x^\prime}
\newcommand{\yp}{y^\prime}
\newcommand{\Uhat}{\widehat{U}}
\newcommand{\Up}{U^\prime}
\newcommand{\Upp}{U^{\prime\prime}}
\newcommand{\Vp}{V^\prime}
\newcommand{\Vhat}{\widehat{V}}
\newcommand{\Vbar}{\overline{V}}
\newcommand{\Ap}{{A^\prime}}
\newcommand{\App}{{A^{\prime\prime}}}
\newcommand{\Cp}{C^\prime}
\newcommand{\Fp}{F^\prime}
\newcommand{\Gp}{G^\prime}
\newcommand{\Fpp}{F^{\prime\prime}}
\newcommand{\Zf}{{\Bbb{Z}}}
\newcommand{\Qf}{{\Bbb{Q}}}
\newcommand{\Rf}{{\Bbb{R}}}
\newcommand{\Cf}{{\Bbb{C}}}
\newcommand{\qacc}{q^{acc}}
\newcommand{\qrej}{q^{rej}}
\newcommand{\qnon}{q^{non}}
\newcommand{\Qadd}{Q_{add}}
\newcommand{\Qacc}{Q_{acc}}
\newcommand{\Qrej}{Q_{rej}}
\newcommand{\Qnon}{Q_{non}}
\newcommand{\Qhalt}{Q_{halt}}
\newcommand{\Qjunk}{Q_{junk}}
\newcommand{\Qaccp}{{Q_{acc}^\prime}}
\newcommand{\Qrejp}{{Q_{rej}^\prime}}
\newcommand{\Qnonp}{{Q_{non}^\prime}}
\newcommand{\Qhaltp}{{Q_{halt}^\prime}}
\newcommand{\Qjunkp}{{Q_{junk}^\prime}}
\newcommand{\Qaccpp}{{Q_{acc}^{\prime\prime}}}
\newcommand{\Qrejpp}{{Q_{rej}^{\prime\prime}}}
\newcommand{\Qnonpp}{{Q_{non}^{\prime\prime}}}
\newcommand{\Qjunkpp}{{Q_{junk}^{\prime\prime}}}
\newcommand{\Cacc}{C_{acc}}
\newcommand{\Crej}{C_{rej}}
\newcommand{\Cnon}{C_{non}}
\newcommand{\Eacc}{E_{acc}}
\newcommand{\Erej}{E_{rej}}
\newcommand{\Enon}{E_{non}}
\def\cent{{\hbox{\rm\rlap/c}}}
\newcommand{\centp}{{\cent}^\prime}
\newcommand{\Bra}[1]{{\langle{#1}|}}
\newcommand{\Ket}[1]{{|{#1}\rangle}}
\newcommand{\BraKet}[2]{{\langle{#1}|{#2}\rangle}}
\newtheorem{theorem}{{\bf Theorem}}[\mytheoremcounter]
\newtheorem{lemma}[theorem]{{\bf Lemma}}
\newtheorem{claim}[theorem]{{\bf Claim}}
\newtheorem{example}[theorem]{{\bf Example}}
\newtheorem{question}[theorem]{{\bf Question}}
\newtheorem{answer}[theorem]{{\bf Answer}}
\newtheorem{conjecture}[theorem]{{\bf Conjecture}}
\newtheorem{proposition}[theorem]{{\bf Proposition}}
\newtheorem{corollary}[theorem]{{\bf Corollary}}
\newtheorem{fact}[theorem]{{\bf Fact}}
\newtheorem{definition}[theorem]{{\bf Definition}}
\newtheorem{remark}[theorem]{{\bf Remark}}
\newtheorem{thoughts}[theorem]{{\bf Thoughts}}
\newenvironment{proof}{ \begin{trivlist} 
                        \item \vspace{-\topsep} \noindent{\bf Proof:}\ }
                      {\rule{5pt}{5pt}\end{trivlist}}
\newcommand{\Case}[2]{\noindent{\bf Case #1:}#2}
\newcommand{\Subcase}[2]{\noindent{\bf Subcase #1:}#2}
\newcommand{\Half}{\frac{1}{2}}
\newcommand{\RtHalf}{\frac{1}{\sqrt{2}}}
\newcommand{\cA}{{\mathcal{A}}}
\newcommand{\cC}{{\mathcal{C}}}
\newcommand{\cE}{{\mathcal{E}}}
\newcommand{\cF}{{\mathcal{F}}}
\newcommand{\cH}{{\mathcal{H}}}
\newcommand{\cI}{{\mathcal{I}}}
\newcommand{\cK}{{\mathcal{K}}}
\newcommand{\cL}{{\mathcal{L}}}
\newcommand{\cO}{{\mathcal{O}}}
\newcommand{\cP}{{\mathcal{P}}}
\newcommand{\cR}{{\mathcal{R}}}
\newcommand{\cS}{{\mathcal{S}}}
\newcommand{\cU}{{\mathcal{U}}}
\newcommand{\Span}{{\mathit{Span}}}
\newcommand{\Ch}[2]{{#1 \choose #2}}
\newcommand{\Ul}[1]{{\underline{#1}}}
\newcommand{\Floor}[1]{{\lfloor #1 \rfloor}}
\newcommand{\ignore}[1]{}
\newcommand{\noignore}[1]{#1}

\newcommand{\RMO}{\mathbf{RMO}}
\newcommand{\UMO}{\mathbf{UMO}}
\newcommand{\RMOe}{\mathbf{RMO}_\epsilon}
\newcommand{\RMM}{\mathbf{RMM}}
\newcommand{\UMM}{\mathbf{UMM}}
\newcommand{\RMMe}{\mathbf{RMM}_\epsilon}

\newcommand{\MOQFA}{\mathbf{MOQFA}}
\newcommand{\MOQFAe}{\mathbf{MOQFA}_\epsilon}
\newcommand{\MMQFA}{\mathbf{MMQFA}}
\newcommand{\MMQFAe}{\mathbf{MMQFA}_\epsilon}
\newcommand{\GQFA}{\mathbf{GQFA}}
\newcommand{\GQFAe}{\mathbf{GQFA}_\epsilon}

\newcommand{\REG}{\mathbf{REG}}
\newcommand{\PFA}{\mathbf{PFA}}
\newcommand{\PFAe}{\mathbf{PFA}_\epsilon}
\newcommand{\GFA}{\mathbf{GFA}}

\newcommand{\Foreach}[2]{\\{\bf\tt{for\ each}} $#1$ {\bf\tt{do}}\+ #2
\- \\ {\bf\tt{rof}}}
\newcommand{\Forloop}[2]{\\{\bf\tt{for}} $#1$ {\bf\tt{do}}\+ #2
\- \\ {\bf\tt{rof}}}
\newcommand{\Ifthen}[2]{\\{\bf\tt{if}} $#1$ {\bf\tt{then}}\+ #2
\- \\ {\bf\tt{fi}}}
\newcommand{\Ifelse}[3]{\\{\bf\tt{if}} $#1$ {\bf\tt{then}}\+ #2
\- \\ {\bf\tt{else}}\+ #3 \- \\ {\bf\tt{fi}}}
\newcommand{\Stmt}[1]{\\$#1$;}
\newcommand{\StartStmt}[1]{\+\kill$#1$;}
\newenvironment{pseudocode}{\begin{tabbing} 
\ \ \ \ \=\ \ \ \ \=\ \ \ \ \=\ \ \ \ \=\ \ \ \ \=\ \ \ \ \=\ \ \ \ \=\
\ \ \ \= } {\end{tabbing}}

\providecommand{\SaveProof}[3]{#3}
\providecommand{\SketchProof}[4]{#3}
\providecommand{\AppendixProof}[3]{}

\newcommand{\MoveProofsToAppendix}{\include{movemacs}}

\newcommand{\ShortSep}{\\ & &}
\newcommand{\LongSep}{}
\providecommand{\DefSep}{\LongSep}
\newcommand{\UseShortSep}{\renewcommand{\DefSep}{\ShortSep}}

\newcommand{\UseAbstract}[2]{#1}
\newcommand{\UseOtherAbstract}{\include{absselect}}

\newcommand{\StretchPage}{ \addtolength{\textheight}{0.05\textheight}
                           \addtolength{\topmargin}{-0.03\textheight}
                         }


\newcommand{\DoFigure}[4]{
                          \begin{figure}[ht]
                            \begin{center}
                              \ \psfig{file=#1,width=#2}\ 
                            \end{center}
                            \caption{#3\label{#4}}
                          \end{figure}
                         }

\providecommand{\captionfontstyle}{}
\def\captionstyle{\captionfontstyle}
\def\boxcaptionstyle{\captionfontstyle\raggedright}

\def\maxfigfraction{.6}

\newdimen\figboxmargin
\figboxmargin = .15in

\newdimen\figboxhang
\figboxhang = 0pt

\def\DVIscaling{1}
\def\globalscaling{1}
\def\figuredirectory{.}
\let\boxer=\llboxer

\def\missingfigure#1{\hbox{Missing figure #1.ps}}

%
%
\catcode `\@=11

\newbox\figurebox

\def\figbox{
\@ifnextchar[{\figboxaux}{\figboxaux[htb]}}

%
%
\long\def\figboxaux[#1#2]#3#4#5#6{
\writepict{{#3}{#4}{#5}{#6}}
\setbox\figurebox\hbox{#3}%
\if l#1\tryleftbox{#4}{#5}{#6}%
\else
\if r#1\tryrightbox{#4}{#5}{#6}%
\else
\if *#1\checktwocoloptions#2]{\box\figurebox}{#4*}{#5}{#6}%
\else\tryonecol[#1#2]{#4}{#5}{#6}%
\fi
\fi
\fi\ignorespaces}

%
%

\long\def\tryleftbox#1#2#3{
\ifdim\wd\figurebox>\maxfigfraction\columnwidth \tryonecol[htb]{#1}{#2}{#3}%
\else\leftbox{\captionbox{\box\figurebox}{#1}{#2}{#3}}\fi}

\long\def\tryrightbox#1#2#3{
\ifdim\wd\figurebox>\maxfigfraction\columnwidth \tryonecol[htb]{#1}{#2}{#3}%
\else\rightbox{\captionbox{\box\figurebox}{#1}{#2}{#3}}\fi}

\def\checktwocoloptions{
\@ifnextchar]{\floatbox[htb}{\floatbox[}}

\long\def\tryonecol[#1]#2#3#4{
\ifdim\wd\figurebox>\columnwidth \floatbox[#1]{\box\figurebox}{#2*}{#3}{#4}%
\else\floatbox[#1]{\box\figurebox}{#2}{#3}{#4}\fi}

%
%
%

\long\def\floatbox[#1]#2#3#4#5{%
\begin{#3}[#1]
\hbox to \hsize{\hfil#2\hfil}
\captionandlabel{#3}{#4}{#5}
\end{#3}
}

%
%
%
%
%
%
%

\long\def\captionbox#1#2#3#4{
\setbox\figurebox\hbox{#1}%
\parbox[t]{\wd\figurebox}{%
\bigskip\box\figurebox
\let\captionstyle=\boxcaptionstyle
\captionandlabel{#2}{#3}{#4}
\bigskip
}}

\def\captionandlabel#1#2#3{
\def\testit{#3}%
\ifx\testit\empty\else
\writecapt{{#1}{#2}{#3}}
\captypeunstarred#1*.
\getcaption#3\endc@ption
\def\testit{#2}
\ifx\testit\empty\else\label{#2}\fi
\fi}

\def\captypeunstarred#1*#2.{
\def\@captype{#1}}

%
%

\def\getcaption{\@ifnextchar[{\getcaptwo}{\getcapone}}
\long\def\getcapone#1\endc@ption{\caption[#1]%
{\def\baselinestretch{1}\Large\normalsize\captionstyle\ignorespaces #1}}
\long\def\getcaptwo[#1]#2\endc@ption{\caption[#1]%
{\def\baselinestretch{1}\Large\normalsize\captionstyle\ignorespaces #2}}

%
%
%

\newdimen\figboxht
\newcount\figboxn

\newcount\figboxlines
\newdimen\figboxwid

\newif\ifisleftbox

\long\def\leftbox#1{%
\setbox\figurebox\hbox{#1}\global\isleftboxtrue
\startmarginbox
\vadjust{\smash{\rlap{\hskip\hsize\hskip\figboxhang
\llap{\raise.7\baselineskip\box\figurebox\hskip\rightskip}}}}%
\endmarginbox%
}

\long\def\rightbox#1{%
\setbox\figurebox\hbox{#1}\global\isleftboxfalse
\startmarginbox
\smash{\llap{\raise.7\baselineskip\box\figurebox\hskip\figboxmargin}}%
\endmarginbox%
}

\def\startmarginbox{%
\ifvmode\passpict\let\endmarginbox=\indent
\else\message{WARNING: marginbox in not in vmode}\hfilneg\ \passpict
\let\endmarginbox=\relax\fi
\figboxht=\dp\figurebox
\advance\figboxht by 1.3\baselineskip
\vskip.95\figboxht\penalty-300\vskip-.95\figboxht
\divide\figboxht by\baselineskip
\global\figboxlines=\figboxht
\global\figboxwid=\wd\figurebox
\global\advance\figboxwid by \figboxmargin
\global\advance\figboxwid by -\figboxhang
\setmypar\noindent}
%

%
%
%
\def\addlines#1{\global\advance\figboxlines by #1\myparshape}
\def\zerolines{\origpar\global\figboxlines=0\myparshape}

%
%
\def\passpict{\par\ifnum\figboxlines>1\vskip\figboxlines\baselineskip
\zerolines\fi}

\def\emptybox#1#2{\hbox to #1{\vbox to #2{\vss}\hss}}


\global\let\origpar=\@@par
\global\let\dopar=\origpar
\global\def\@@par{\dopar}
\@setpar{\dopar}

\def\setmypar{\global\let\dopar=\mypar
\global\prevgraf=0\myparshape}

\def\mypar{\origpar\global\advance\figboxlines by -\prevgraf%
\global\prevgraf=0\myparshape}

\def\myparshape{\relax%
\ifnum\figboxlines>1\theparshape \else
\global\hangindent=0pt\global\hangafter=1
\global\let\dopar=\origpar\fi}

\def\theparshape{%
\ifisleftbox\global\hangindent=-\figboxwid 
\else\global\hangindent=\figboxwid \fi
\global\hangafter=-\figboxlines \global\advance\hangafter by 1%
}

%
%
%
%

\def\definefnum#1{
\def\fnum@figure{Figure \ref{#1}}%
\def\fnum@table{Table \ref{#1}}%
\def\fnum@code{Algorithm \ref{#1}}%
}

\def\writepict#1{}
\def\writecapt#1{}

%

\def\journalpicts#1{
\newwrite\pictfile
\newwrite\captfile
\openout\pictfile\jobname.pic
\openout\captfile\jobname.cap
\gdef\writepict##1{\unexpandedwrite\pictfile{\doit##1}}%
\gdef\writecapt##1{\unexpandedwrite\captfile{\doit##1}}%
\global\let\ENDdocument=\enddocument
\gdef\enddocument{\DOjournalpicts{#1}\ENDdocument}
}

%
%
\def\DOjournalpicts#1{{%
\def\writepict##1{}\closeout\pictfile
\def\writecapt##1{}\closeout\captfile
\@fileswfalse
\onecolumn
\def\globalscaling{#1}
\def\doit##1##2##3##4{
\figboxaux[t]{\hss##1\hss}{##2}{}{}%
\vspace*{1in}
\definefnum{##3}
\captionandlabel{##2}{}{##4}
\clearpage}%
\input main2.pic
\def\doit##1##2##3{
\definefnum{##2}
\captionandlabel{##1}{}{##3}}%
\raggedright\let\captionstyle=\raggedright
\def\@makecaption##1##2{##1: ##2\par}
\input\jobname.cap
}}

%
%
%
%
%
%

%
%

\def\llboxer#1{\vbox to \figboxht{\vfil\hbox to \figboxwid{#1\hfill}}}
\def\lcboxer#1{\vbox to \figboxht{\vfil\hbox to \figboxwid{\hfill#1\hfill}}}
\def\oldboxer#1{\vbox to \figboxht{\vfil
                      \hbox to \figboxwid{\hfill\llap{#1\hskip4.25in}\hfill}}}
\def\ulboxer#1{\vbox to \figboxht{\hbox to \figboxwid{#1\hfill}\vfil}}
\def\ccboxer#1{\vbox to \figboxht{\vfil
                        \hbox to \figboxwid{\hfill#1\hfill}\vfil}}
\def\basicboxer#1{\vbox to \figboxht{\vfil\hbox to \figboxwid{#1}\vskip-.1in}}

%
{\catcode`\p=12\catcode`\t=12
\gdef\removedimen#1pt{#1}}

\def\defscaled#1#2{#2=\DVIscaling#2%
\xdef#1{\expandafter\removedimen\the#2}}

\def\DVIspace{ }
\newdimen\hscalefactor
\newdimen\vscalefactor

\def\scale#1{\horizscale{#1}\vertscale{#1}}
\def\horizscale#1{\hscalefactor=#1\hscalefactor\figboxht=#1\figboxht}
\def\vertscale#1{\vscalefactor=#1\vscalefactor\figboxwid=#1\figboxwid}

%

\def\boxps{%
\@ifnextchar[{\boxpsaux}{\boxpsaux[\relax]}}

\def\boxpsaux[#1]#2#3#4#5{%
{\figboxwid#4\figboxht#5\hscalefactor=1pt\vscalefactor=1pt%
\scale{#3}%
\scale{\globalscaling}%
#1%
\defscaled\DVIhscale\hscalefactor%
\defscaled\DVIvscale\vscalefactor%
\boxer{\includegraphics{\figuredirectory/#2}}}%
}

%
%
%
\newread\Epsffilein
\newif\ifEpsffileok
\newif\ifEpsfbbfound

\newdimen\pspoints
\pspoints=1in
\divide\pspoints by 72

\def\boxeps{%
\@ifnextchar[{\boxepsaux}{\boxepsaux[\relax]}}
\def\boxepsaux[#1]#2#3{%
%
%
\openin\Epsffilein=\figuredirectory/#2 
\ifeof\Epsffilein\message{I couldn't open \figuredirectory/#2 }%
\missingfigure{#2}
\else
%
%
   {\Epsffileoktrue\Epsfbbfoundfalse
    \catcode`\%=11 \catcode`\\=11
    \catcode`\{=11 \catcode`\}=11
    \catcode`\$=11 \catcode`\^=11
    \catcode`\&=11 \catcode`\#=11
    \catcode`\~=11 \catcode`\_=11
    \loop
       \read\Epsffilein to \Epsffileline
       \ifeof\Epsffilein\Epsffileokfalse\else
%
%
          \expandafter\Epsfaux\Epsffileline . .\\%
       \fi
   \ifEpsffileok\repeat
   \ifEpsfbbfound
        \figboxht=\Epsfury\pspoints
        \advance\figboxht by-\Epsflly\pspoints
        \figboxwid=\Epsfurx\pspoints
        \advance\figboxwid by-\Epsfllx\pspoints
   \else
        \message{No bounding box comment in \figuredirectory/#2 }%
        \figboxwid=2in\figboxht=1in%
   \fi%
   \immediate\closein\Epsffilein
   \hscalefactor=1pt\vscalefactor=1pt%
   \scale{#3}%
   \scale{\globalscaling}%
   #1%
   \defscaled\DVIhscale\hscalefactor
   \defscaled\DVIvscale\vscalefactor
   \hscalefactor=-\Epsfllx\hscalefactor
   \hscalefactor=1.00375\hscalefactor
   \defscaled\DVIhoffset\hscalefactor
   \vscalefactor=-\Epsflly\vscalefactor
   \vscalefactor=1.00375\vscalefactor
   \defscaled\DVIvoffset\vscalefactor
   \llboxer{\includegraphics{\figuredirectory/#2}} }%
\fi
}%
%
%
{\catcode`\%=11 \global\let\Epsfpar=\par
\global\let\Epsfpercent=
%
%
\long\def\Epsfaux#1#2 #3\\{\relax\ifx#1\Epsfpercent
   \def\testit{#2}\ifx\testit\Epsfbblit
      \Epsfsize #3 . . . .\\%
      \global\Epsffileokfalse
      \global\Epsfbbfoundtrue
   \fi\else\ifx#1\Epsfpar\else\global\Epsffileokfalse\fi\fi}%
%
%
\def\Epsfsize#1 #2 #3 #4 #5\\{\global\def\Epsfllx{#1}\global\def\Epsflly{#2}%
   \global\def\Epsfurx{#3}\global\def\Epsfury{#4}}%

\catcode`\@=12                  

%
%
%
%

\def\pic#1;#2;#3;#4\par{\picsc#1;#2;#3;1;#4\par}

\def\picsc#1;#2;#3;#4;#5\par{
\figbox[htb]{\boxeps{#1}{#4}
}{figure}{#1}{%
#5}}

\def\mpic#1;#2;#3;#4\par{\mpicsc#1;#2;#3;1;#4\par}

\def\mpicsc#1;#2;#3;#4;#5\par{
\figbox[l]{\boxeps{#1}{#4}
}{figure}{#1}{%
#5}}

\bibliographystyle{alpha}

\title{The Rectilinear Crossing Number of $K_{10}$ is $62$}
\author{ Alex Brodsky\thanks{Supported by NSERC PGSB}\\
       \and 
         Stephane Durocher\\
       \and Ellen Gethner\thanks{{\tt \{abrodsky,durocher,egethner\}@cs.ubc.ca},
Department of Computer Science, 
University of British Columbia, 201 - 2366 Main Mall,
Vancouver, B.C., Canada,
V6T 1Z4}
        }

\maketitle

\begin{center}
\begin{raggedright}
{\it ``Oh what a tangled web we weave...''}\\
Sir Walter Scott
\end{raggedright}
\end{center}

\abstract{A drawing of a graph $G$ in the plane is said to be a
rectilinear drawing of $G$ if the edges are required to be line
segments (as opposed to Jordan curves). We assume no three vertices are
collinear.  The rectilinear crossing number of $G$ is the fewest number
of edge crossings attainable over all rectilinear drawings of $G$.
Thanks to Richard Guy, exact values of the rectilinear crossing number
of $K_n$, the complete graph on $n$ vertices, for $n = 3,\ldots,9$, are
known \cite{Gu72,WhBe78,Fi00,SlA014540}. Since 1971, thanks to the
work of David Singer \cite{Si71,Ga86}, the rectilinear crossing number
of $K_{10}$ has been known to be either 61 or 62, a deceptively
innocent and tantalizing statement. The difficulty of determining the
correct value is evidenced by the fact that Singer's result has
withstood the test of time. In this paper we use a purely combinatorial
argument to show that the rectilinear crossing number of $K_{10}$ is
62. Moreover, using this result, we improve an asymptotic lower bound
for a related problem. Finally, we close with some new and old open
questions that were provoked, in part, by the results of this paper,
and by the tangled history of the problem itself.}

\section{Introduction and History} 
Mathematicians and Computer Scientists are well acquainted with the
vast sea of crossing number problems, whose 1944 origin lies in a scene
described by Paul Tur{\'a}n. The following delightful excerpt, taken
from \cite{Gu69}, has appeared numerous times in the literature over
the years, and is now known as ``Tur{\'a}n's brick factory problem.''

\begin{quote}
{\it {\em[sic.]}In 1944 our labor cambattation had the extreme luck to
work---thanks to some very rich comrades---in a brick factory near
Budapest. Our work was to bring out bricks from the ovens where they
were made and carry them on small vehicles which run on rails in some
of several open stores which happened to be empty. Since one could
never be sure which store will be available, each oven was connected by
rail with each store.  Since we had to settle a fixed amount of loaded
cars daily it was our interest to finish it as soon as possible. After
being loaded in the (rather warm) ovens the vehicles run smoothly with
not much effort; the only trouble arose at the crossing of two rails.
Here the cars jumped out, the bricks fell down; a lot of extra work and
loss of time arose. Having this experience a number of times it
occurred to me why on earth did they build the rail system so
uneconomically; minimizing the number of crossings the production could
be made much more economical.}
\end{quote}

\noindent And thus the crossing number of a graph was born. The
original concept of the crossing number of the complete bipartite graph
$K_{m,n}$, as inspired by the previous quotation, was addressed by
K{\"o}vari, S{\'o}s, and Tur{\'a}n in \cite{KoSoTu54}.  Following suit,
Guy \cite{Gu60} initiated the hunt for the crossing number of $K_n$.\\

\noindent Precisely,

\begin{definition}
Let {\rm G} be a graph drawn in the plane such that the edges of
{\rm G} are Jordan curves, no three vertices are collinear, no vertex
is contained in the interior of any edge, and no three edges may
intersect in a point, unless the point is a vertex. The \myem{crossing
number of G}, denoted $\bf\mathbf{cr}(G)$, is the minimum number of
edge crossings attainable over all drawings of {\rm G} in the plane.
A drawing of {\rm G} that achieves the minimum number of edges
crossings is called \myem{optimal}.  
\end{definition}

In this paper we are interested in drawings of graphs in the plane in
which the edges are line segments.

\begin{definition}
Let {\rm G} be a graph drawn in the plane with the requirement that
the edges are line segments, no three vertices are collinear, and no
three edges may intersect in a point, unless the point is a vertex.
Such a drawing is said to be a \myem{rectilinear drawing of G}.
The \myem{rectilinear crossing number of G}, denoted ${\bf
\overline{\mathbf{cr}}(G)}$, is the fewest number of edge crossings
attainable over all rectilinear drawings of {\rm G}. Any such a drawing
is called \myem{optimal}.  
\end{definition}

\subsection{A Few General Results}
We mention a small variety of papers on crossing numbers problems for
graphs drawn in the plane that merely hint at the proliferation of
available (and unavailable!) results. Other important results will be
highlighted in Section \ref{futurework}.

Garey and Johnson \cite{GaJo83} showed that the problem of determining
the crossing number of an arbitrary graph is NP-complete. Leighton
\cite{Le84} gave an application to VLSI design by demonstrating a
relationship between the area required to design a chip whose circuit
is given by the graph $G$ and the rectilinear crossing number of $G$.
Bienstock and Dean \cite{BiDe93} produced an infinite family of graphs
$\{G_m\}$ with $\regcr{G_m} = 4$ for every $m$ but for which
$sup_{m}\{\reccr{G_m}\} = \infty.$ Kleitman \cite{Kl70,Kl76} completed
the very difficult task of determining the exact value of
$\regcr{K_{5,n}}$ for any $n\in {\Zf}^+$.  Finally, a crucial method of
attack for both rectilinear crossing number and crossing number
problems has been that of determining the parity (i.e., whether the
crossing number is even or odd). See, for example,
\cite{Ha76,Kl70,Kl76,ArRi88, HaTh96}.

Crossing number problems are inherently rich and numerous, and have
captured the attention of a diverse community of researchers. For a
nice exposition of current open questions as well as a plethora of
references, see the recent paper of Pach and T{\'o}th \cite{PaTo00}.

\subsection{Closer to Home: $\reccr{K_n}$}
Many papers, dating back as far as 1954 \cite{KoSoTu54}, have addressed
the specific problem of determining $\regcr{K_{m,n}}$ and
$\regcr{K_n}$.  For a nice overview see Richter and Thomassen
\cite{RiTh97}. For those who are tempted by some of the problems
mentioned in this paper, it is imperative to read \cite{Gu69} for
corrections and retractions in the literature.

Our present interest is that of finding
$\reccr{K_n}$ whose notion was first introduced by Harary and
\figbox[l]{\mbox{ \begin{tabular}{|l|l|}\hline
$K_n$ & $\reccr{K_n}$ \\ \hline
$K_3$    & 0 \\
$K_4$    & 0 \\
$K_5$    & 1 \\
$K_6$    & 3 \\
$K_7$    & 9 \\
$K_8$    & 19 \\
$K_9$    & 36 \\ \hline
$K_{10}$ & {\it 61 or 62} \\ \hline
\end{tabular}}}{table}{tab:small_cr}{$\reccr{K_n}$}
Hill \cite{HaHi62}. As promised in the abstract, the small values of
$\reccr{K_n}$ are known through $n = 9$, which can be found in
\cite{Gu72,WhBe78,Fi00} and \cite[sequence A014540]{SlA014540}; see
Table~\ref{tab:small_cr}.  Ultimately, the $n = 10$ entry
\cite{Si71,Ga86} will be the focus of this paper.

Asymptotics have played an important role in deciphering some of the
mysteries of $\reccr{K_n}$. To this end, it is well known (see for
example \cite{ScWi94}) that
$\lim_{n\rightarrow\infty}\frac{\reccr{K_n}}{\binom{n}{4}}$ exists and
is finite; let

\begin{equation}\label{equ:rectasymptotics}
\overline{\nu}^* = \lim_{n\rightarrow\infty}\frac{\reccr{K_n}}{\binom{n}{4}}.
\end{equation}

H.F. Jensen \cite{Je71} produced a specific rectilinear drawing of
$K_n$ for each $n$, which availed itself of a formula, denoted
$j(n)$, for the exact number of edge crossings. In particular,

\begin{equation}\label{equ:jensenformula}
j(n) = \left\lfloor{\frac{7n^4 - 56n^3 + 128n^2 + 
                                48n\left\lfloor\frac{n - 7}{3}\right\rfloor +
108}{432}
                         }\right\rfloor,
\end{equation}

\noindent from which it follows that $\reccr{K_n} \leq j(n)$ and that
$\overline{\nu}^* \leq .3\overline{8}$. Moreover, it follows from work
in \cite{Si71} as communicated in \cite{Wi97,BrDuGe00} that

\begin{equation}\label{equ:oldasymptotics}
\frac{61}{210} = .290476 \leq \overline{\nu}^* \leq .3846.
\end{equation}

\noindent In Section~\ref{sec:K_9}, for completeness of exposition we
reproduce the argument in \cite{Si71} that $\reccr{K_{10}} > 60$, which
is required to obtain the lower bound in equation
(\ref{equ:oldasymptotics}).

In the recent past, Scheinerman and Wilf \cite{ScWi94,Wi97,Fi00} have
made an elegant connection between $\overline{\nu}^*$ and a variation
on Sylvester's four point problem. In particular, let $R$ be any open
set in the plane with finite Lebesgue measure, and let $q(R)$ be the
probability of choosing four points uniformly and independently at
random in $R$ such that all four points are on a convex hull. Finally,
let $q_*=\inf_R\{q(R)\}$. Then it is shown that $q_*=\overline{\nu}^*.$

Most recently, Brodsky, Durocher, and Gethner \cite{BrDuGe00} have
reduced the upper bound in equation (\ref{equ:oldasymptotics}) to
.3838. In the present paper, as a corollary to our main result, that
$\reccr{K_{10}} = 62$, we increase the lower bound in equation
(\ref{equ:oldasymptotics}) to approximately .30.

\section{Outline of the proof that $\reccr{K_{10}} = 62$}
As mentioned in the abstract, the main purpose of this paper is to
settle the question of whether $\reccr{K_{10}} = 61$ or 62. Our
conclusion, based on a combinatorial proof, is that $\reccr{K_{10}} =
62$. The following statements, which will be verified in the
next sections, constitute an outline of the proof. As might be
expected, given the long history of the problem and its variants,
there are many details of which we must keep careful track.

\figbox[h]{\boxeps{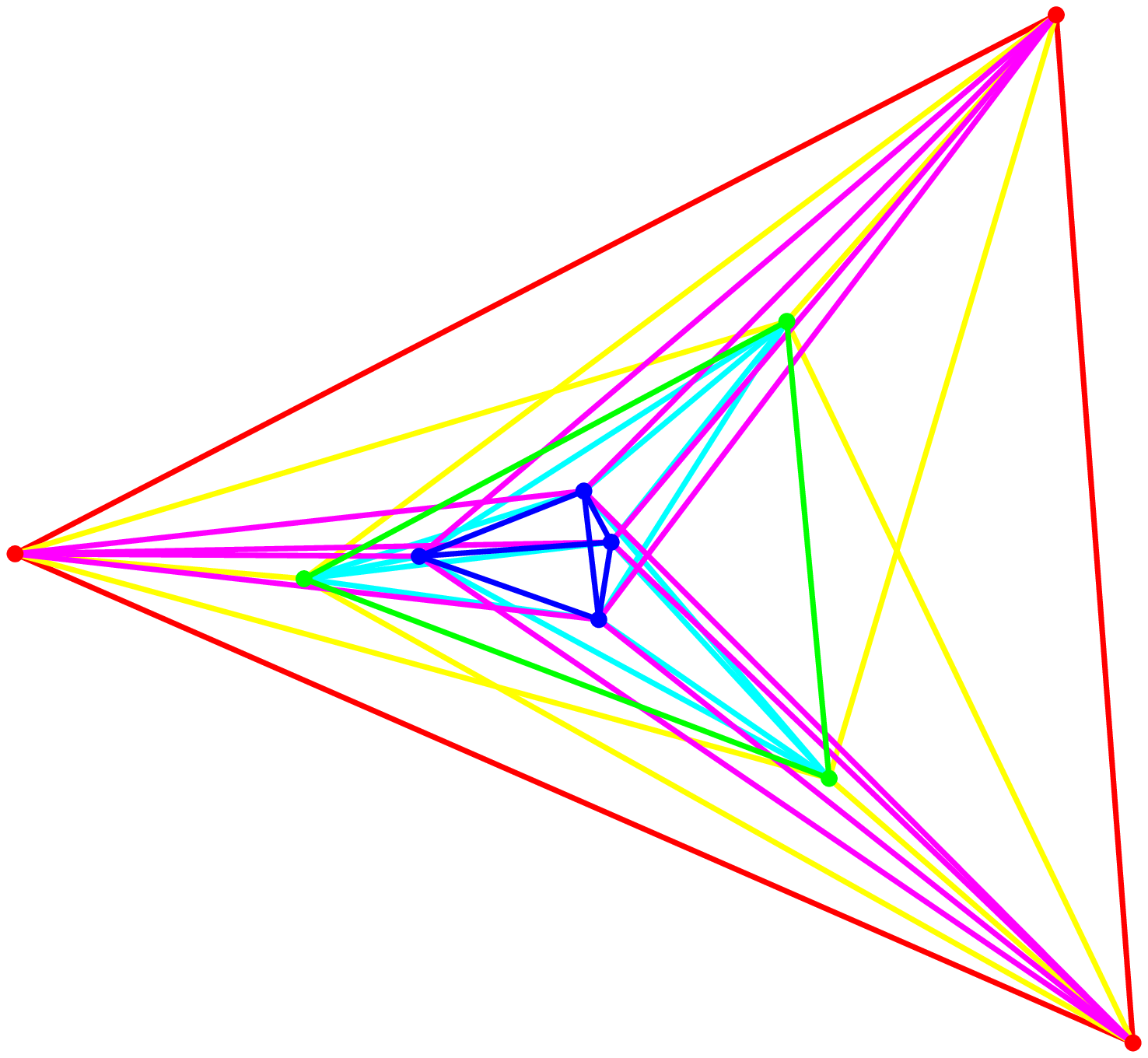}{0.35}}{figure}{fig:colorsinger}{The 
reader is invited to count the number of edge crossings in this optimal 
drawing of $K_{10}$.}

\begin{enumerate}
\item Any optimal rectilinear drawing of $K_9$ consists of three nested
triangles: an outer, middle, and inner triangle.  For purposes of both
mnemonic and combinatorial considerations, we colour the vertices of
the outer triangle {\it red}.  Similarly, the vertices of the middle
triangle will be coloured {\it green} and the vertices of the inner
triangle will be coloured {\it blue}.  For those who are accustomed to
working with computers, the mnemonic is that the vertices of the outer,
middle, and inner triangles correspond to {\it RGB}.

Continuing in this vein, each of the edges of the $K_9$ drawing are
coloured by way of the colour(s) of the two vertices on which they are
incident.  For example, an edge incident on a red vertex and a green
vertex will naturally be coloured yellow. An edge incident on two red
vertices (i.e., an edge of the outer triangle) will be coloured red,
and so on.  This step is done purely for purposes of visualization. For
examples, see Figures \ref{fig:k9_ccc}, \ref{fig:k9_ccv}, and
\ref{fig:k9_cvv}.

Combinatorially, an edge crossing has a label identified by the four
(not necessarily distinct) colours of the two associated edges, 
\Xng{wx}{yz}, where w,x,y,z$\in\{r,g,b\}$.

\item A drawing of $K_{10}$ with 61 crossings must contain a drawing of 
$K_9$ with 36 crossings and must have a convex hull that is a triangle.

\item In any pair of nested triangles with all of the accompanying
edges (i.e., a $K_6$), we exploit a combinatorially invariant: the
subgraph induced by a single outer vertex together with the three
vertices of the inner triangle is a $K_4$. There are exactly two
rectilinear drawings of $K_4$. That is, the convex hull of rectilinear
drawing of $K_4$ is either a triangle or a quadrilateral. If the
former, since the drawing is rectilinear, there are no edge crossings.
If the latter, there is exactly one edge crossing, namely that of the
two inner diagonals.

\item With the above machinery in place, we enumerate the finitely many
cases that naturally arise. In each case we find a lower bound for the
number of edge crossings. In all cases, the result is at least 62.

\item Singer \cite{Si71} produced a rectilinear drawing of $K_{10}$ with 
62 edge crossings, which is exhibited in \cite[p.  142]{Ga86}.  This
together with the work in step 4 implies that $\reccr{K_{10}} = 62$;
see Figure~\ref{fig:colorsinger}.  
\end{enumerate}

The remainder of this paper is devoted to the details of the outline
just given, the improvement of the lower bound in equation
(\ref{equ:oldasymptotics}), and finally, a list of open problems and
future work.

\section{Edge Crossing Toolbox}
\subsection{Definitions}
We assume that all drawings are in general position, i.e., no three
vertices are collinear.  A rectilinear drawing of a graph is
decomposable into a set of convex \myem{hulls}.  The \myem{first hull}
\figbox[r]{\boxeps{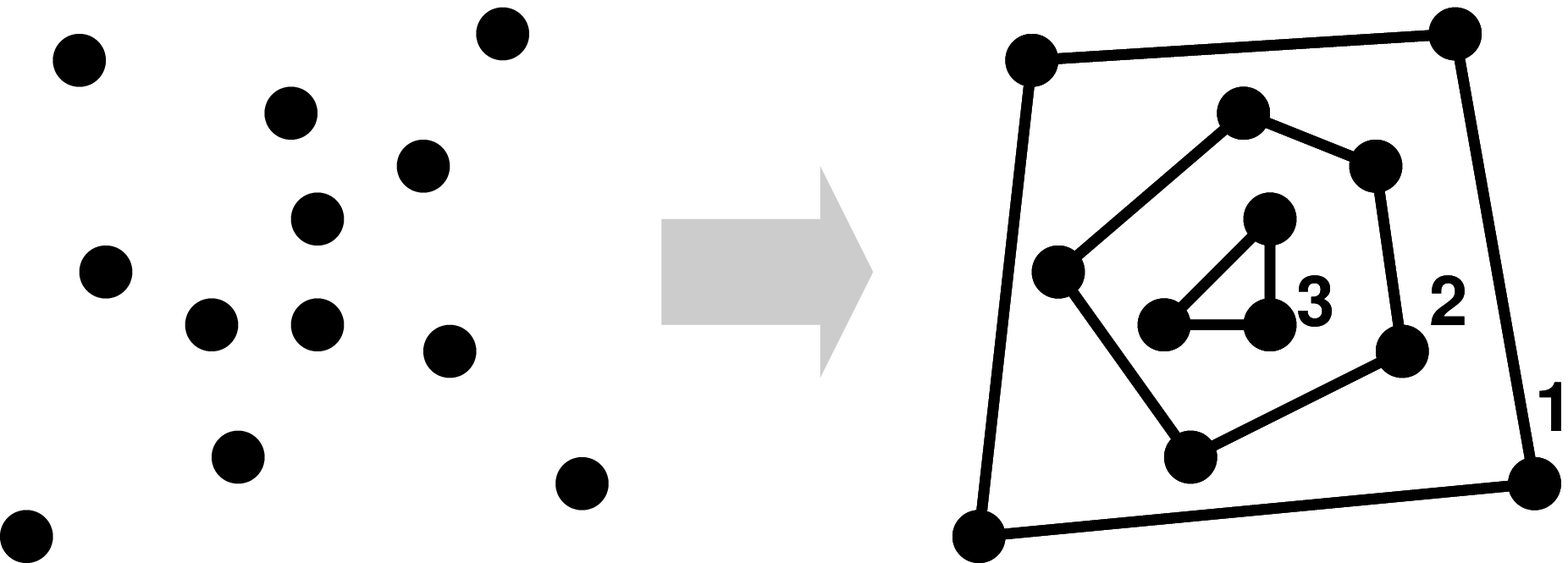}{0.25}}{figure}{fig:convex}{}
of a drawing is the convex hull.  The \myem{$i$th hull} is the
convex hull of the drawing of the subgraph strictly contained within
the $(i-1)$st hull.

The \myem{responsibility} of a vertex in a rectilinear drawing, defined
in~\cite{Gu72}, is the total number of crossings on all edges incident
on the vertex.

\figbox[r]{\boxeps{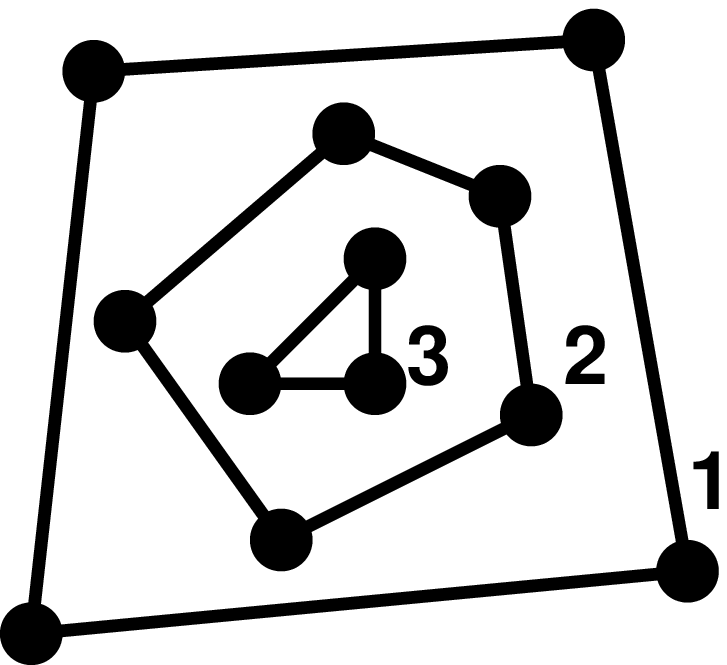}{0.25}}{figure}{fig:nested}{}
A \myem{polygon of size $k$} is a rectilinear drawing of a non-crossing
cycle on $k$ vertices.  A polygon is \myem{contained} within another
polygon if all the vertices of the former are strictly contained within
the boundaries of the latter; the former is termed the \myem{inner}
polygon and the latter, the \myem{outer} polygon.  We say that $n$
polygons are \myem{nested} if the $(i+1)$st polygon is contained within
the $i$th polygon for all $1 \leq i < n$.  A triangle is a polygon of
size three and every hull is a convex polygon.

\figbox[r]{\boxeps{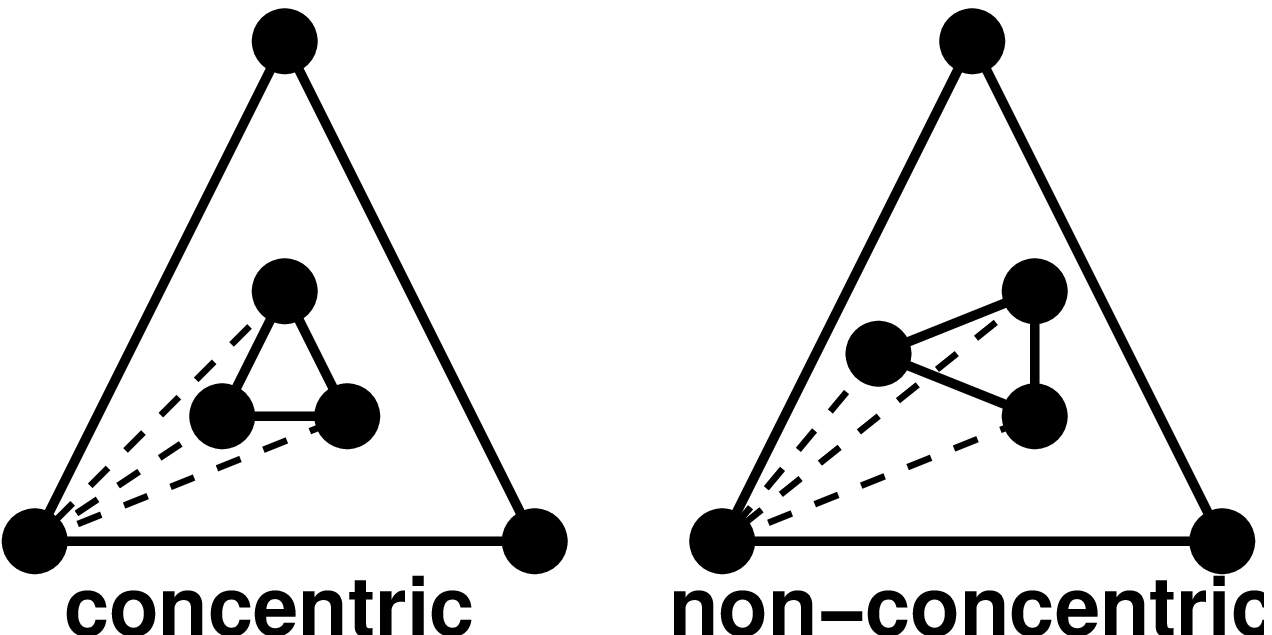}{0.25}}{figure}{fig:concentric}{}
A rectilinear drawing of $K_n$ is called a \myem{nested triangle
drawing} if any pair of hulls of the drawing are nested
triangles.

Two polygons are \myem{concentric} if one polygon contains the other
polygon and any edge between the two polygons intersects neither the
inner nor the outer polygon.  Given two nested polygons, if the inner
polygon is not a triangle then the two polygons a priori cannot be
concentric.  A crossing of two edges is called a \myem{non-concentric
crossing} if one of its edges is on the inner hull and the other has
endpoints on the inner and outer hulls.

We know that the first hull of an optimal rectilinear drawing of
$K_9$ must be a triangle~\cite{Gu72}.  Furthermore, in
Subsection~\ref{sec:K_10} we will reproduce a theorem from \cite{Si71},
that the outer two hulls of a rectilinear drawing of $K_9$ must be
triangles.

\figbox[l]{\boxeps{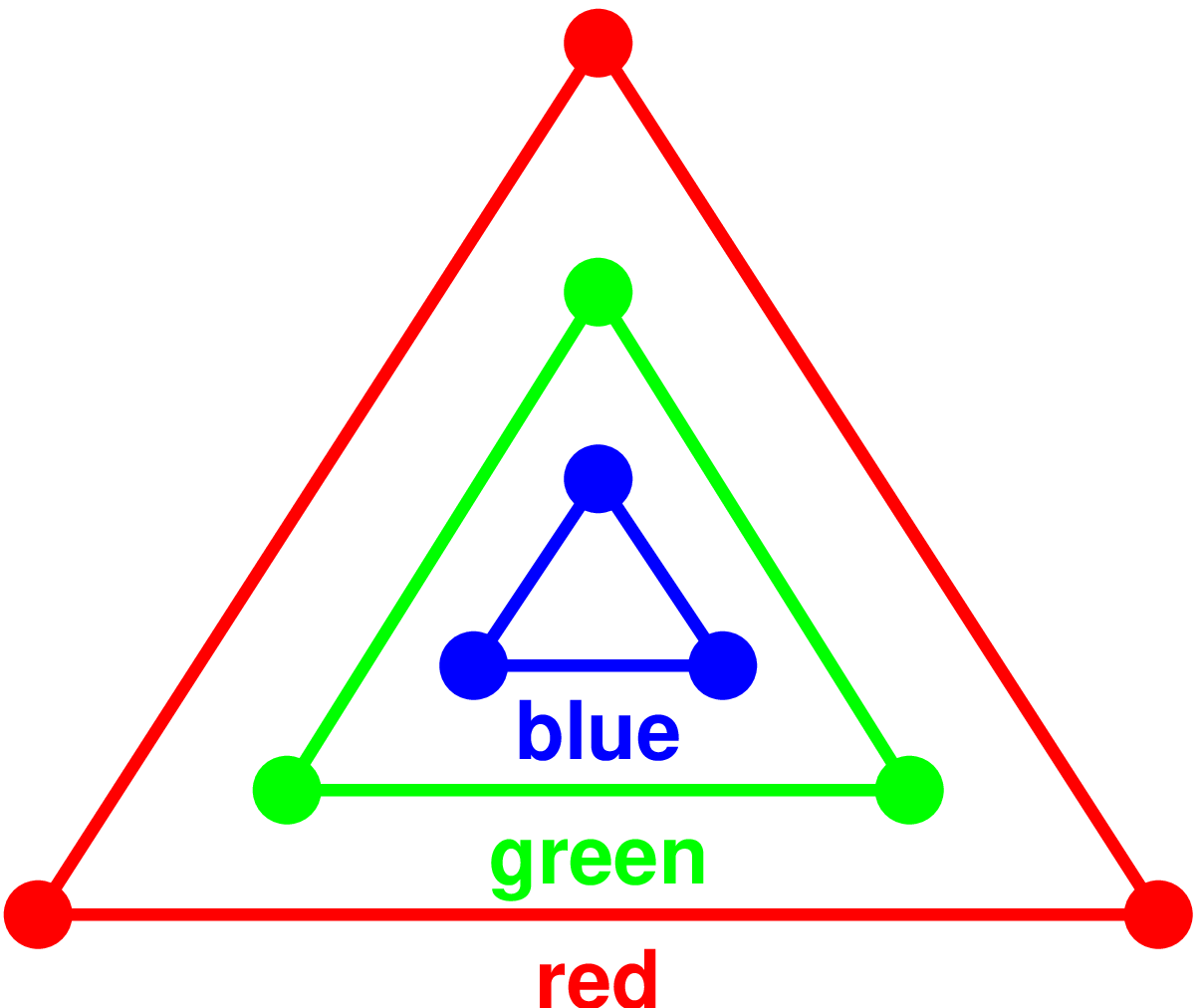}{0.25}}{figure}{fig:colours}{}
For clarity, we colour the outer triangle red, the second triangle
green, and the inner triangle blue.  The vertices of a triangle take on
the same colour as the triangle, and an edge between two vertices is
labeled by a colour pair, e.g., red-blue (rb).  A crossing of two edges
is labeled by the colours of the comprising edges, e.g.,
red-blue$\times$red-green (\Xng{rb}{rg}).  A crossing is called
\myem{2-coloured} if only two colours are involved in the crossing.
This occurs when both edges are incident on the same two triangles,
e.g., \Xng{rg}{rg}, or when one of the edges belongs to the triangle
\figbox[l]{\boxeps{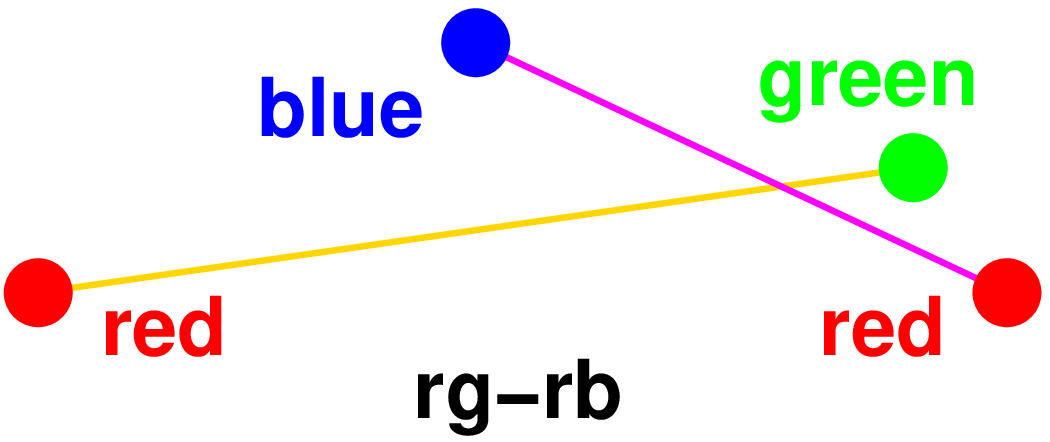}{0.25}}{figure}{fig:crossing}{} 
that the other edge is incident on, e.g., \Xng{rg}{gg}.  A
\myem{3-coloured} crossing is one where the two edges that are involved
are incident on three different triangles, e.g., \Xng{rb}{rg}.
A \myem{4-coloured} crossing is defined similarly.

Crossings may be referred to by their full colour specification, the
colours of an edge comprising the crossing, or the colour of a vertex
comprising the crossing.  For example, an \Xng{rg}{rb}\ crossing is
fully specified by the four colours, two per edge; the crossing is also
a red-blue crossing and a red-green crossing because one of the edges
is coloured red-blue and the other is coloured red-green.  Since the
edges of the crossing are incident on the red, green and blue vertices,
the crossing may also be called red, green or blue; a \Xng{rg}{rg}\ 
crossing is neither red-blue nor blue.

\subsection{Configurations}
\figbox[l]{\boxeps{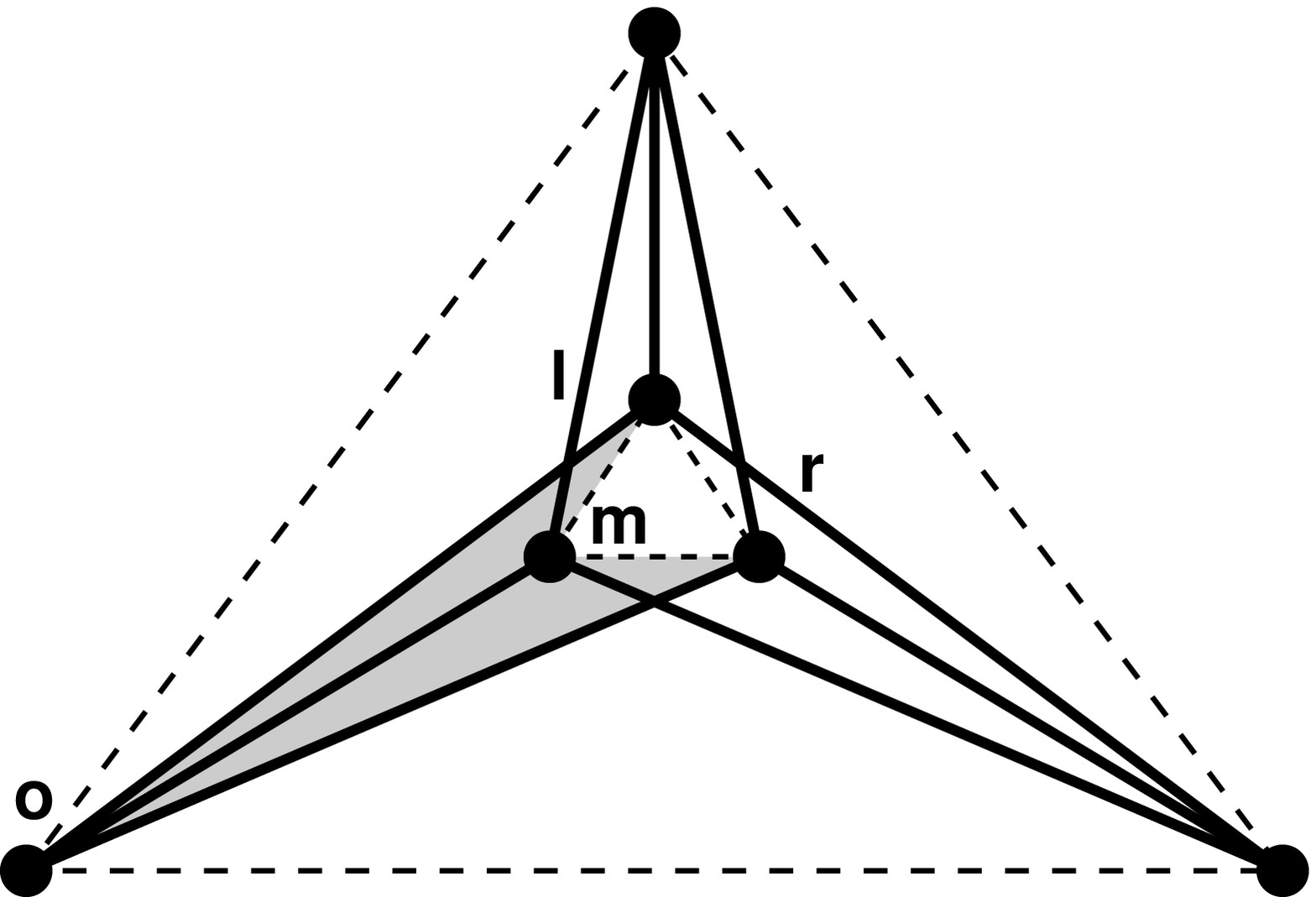}{0.25}}{figure}{fig:ccc}{CCC}
Given a nested triangle drawing of $K_6$, a \myem{kite} is a set of
three edges radiating from a single vertex of the outer triangle to
each of the vertices of the inner triangle.  A kite comprises four
vertices:  the origin vertex, labeled $o$, from which the kite
originates, and three internal vertices.  The internal vertices are
labeled in a clockwise order, with respect to the origin vertex,  by
the labels left ($l$), middle ($m$), and right $(r)$; the angle
$<\!\!lor$ must be acute.  The kite also has three edges, two outer
edges, $(o,l)$ and $(o,r)$, and the inner edge $(o,m)$.  The origin
\figbox[r]{\boxeps{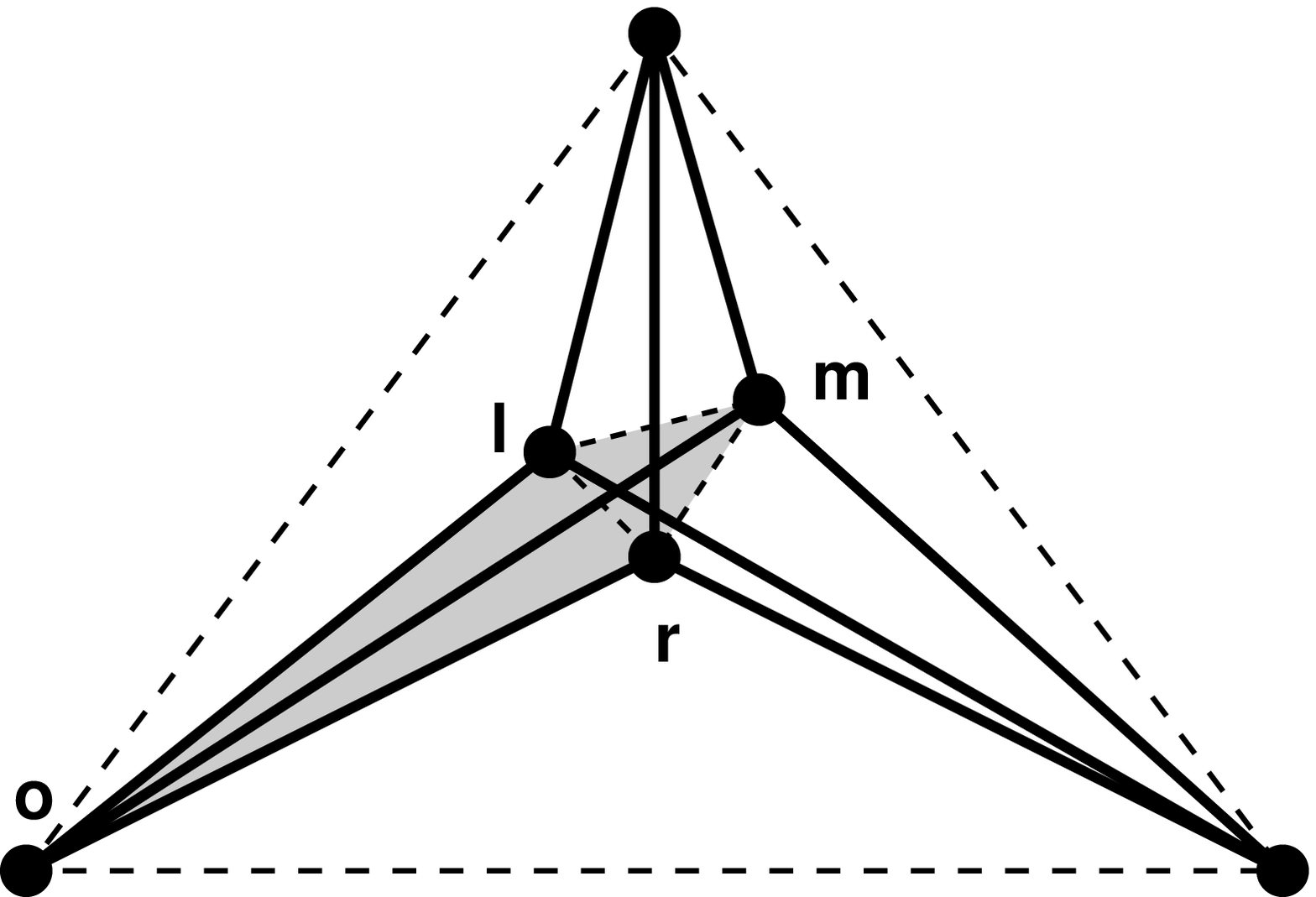}{0.25}}{figure}{fig:vvv}{VVV} 
vertex corresponds to the vertex on the outer triangle and the middle
vertex is located within the sector defined by $<\!\!lor$; see
Figure~\ref{fig:ccc}.  A kite is called \myem{concave} if $m$ is
contained within the triangle $\Delta lor$, see Figure~\ref{fig:ccc},
and is called \myem{convex} if $m$ is not contained in the triangle
$\Delta lor$, see Figure~\ref{fig:vvv}.  We shall denote a convex kite
by V and a concave kite by C.  A vertex is said to be inside a kite if
it is within the convex hull of that kite, otherwise the vertex is said
to be outside the kite.

\figbox[r]{\boxeps{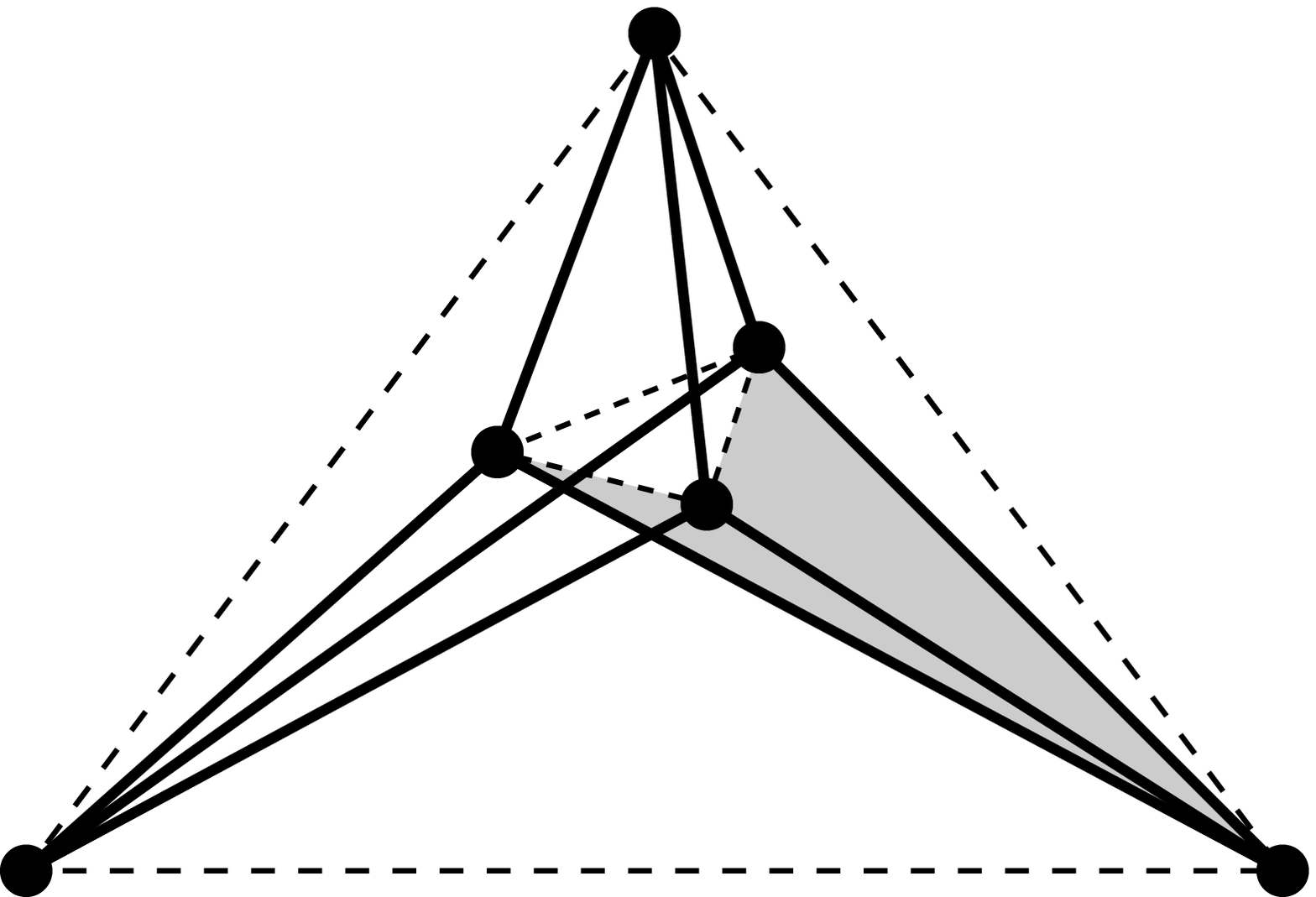}{0.25}}{figure}{fig:cvv_a}{CVV}
A \myem{configuration} of kites is a set of three kites in a nested
triangle drawing of $K_6$.  Each kite originates from a different
vertex of the outer triangle and is incident on the same inner triangle.
There are four different configurations: CCC, CCV, CVV, and VVV,
corresponding to the number of concave and convex kites in the
drawing.

\figbox[r]{\boxeps{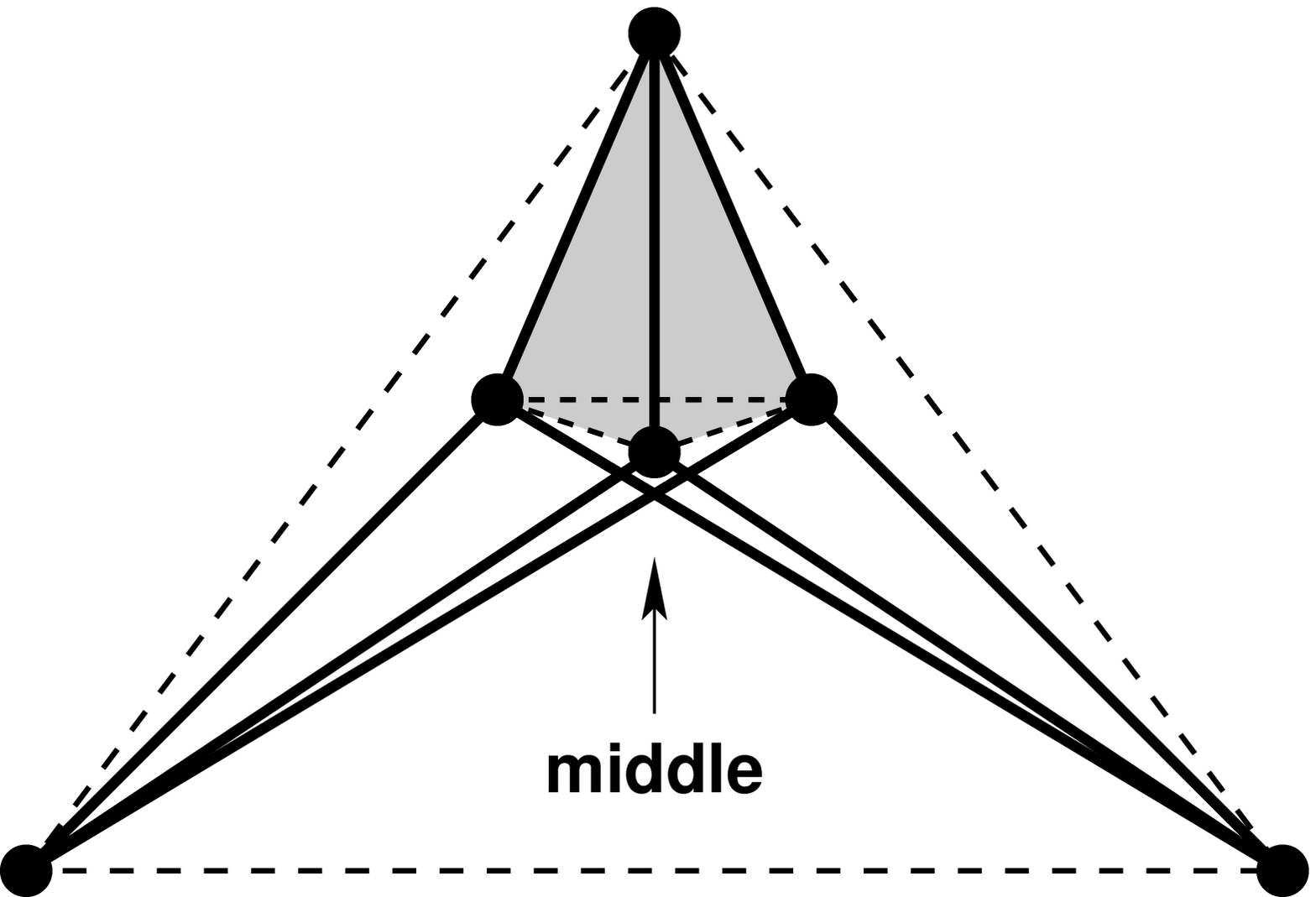}{0.25}}{figure}{fig:ccv_s}{Unary CCV}
A configuration determines how many non-concentric crossings there are,
i.e., the number of edges intersecting the inner triangle; CCC has zero,
CCV has one, CVV has two, and VVV has three non-concentric edge
crossings.  A \myem{sub-configuration} corresponds to the number of
distinct middle vertices of concave kites; this can vary depending on
whether the concave kites share the middle vertex.

\begin{remark}
A CCV configuration is the only one that has more than one
sub-configuration.  A VVV configuration has no concave kites, a
CVV configuration has only one concave kite, and in a CCC
configuration no two kites share a middle vertex.  
\end{remark}

In configuration CCC, Figure~\ref{fig:ccc}, there are three distinct
middle vertices of concave kites, and in configuration VVV,
Figure~\ref{fig:vvv}, there are zero because there are no concave
kites.  Configuration CVV, Figure~\ref{fig:cvv_a}, has only one
middle vertex that belongs to a concave kite because it has only one
concave kite.

\figbox[l]{\boxeps{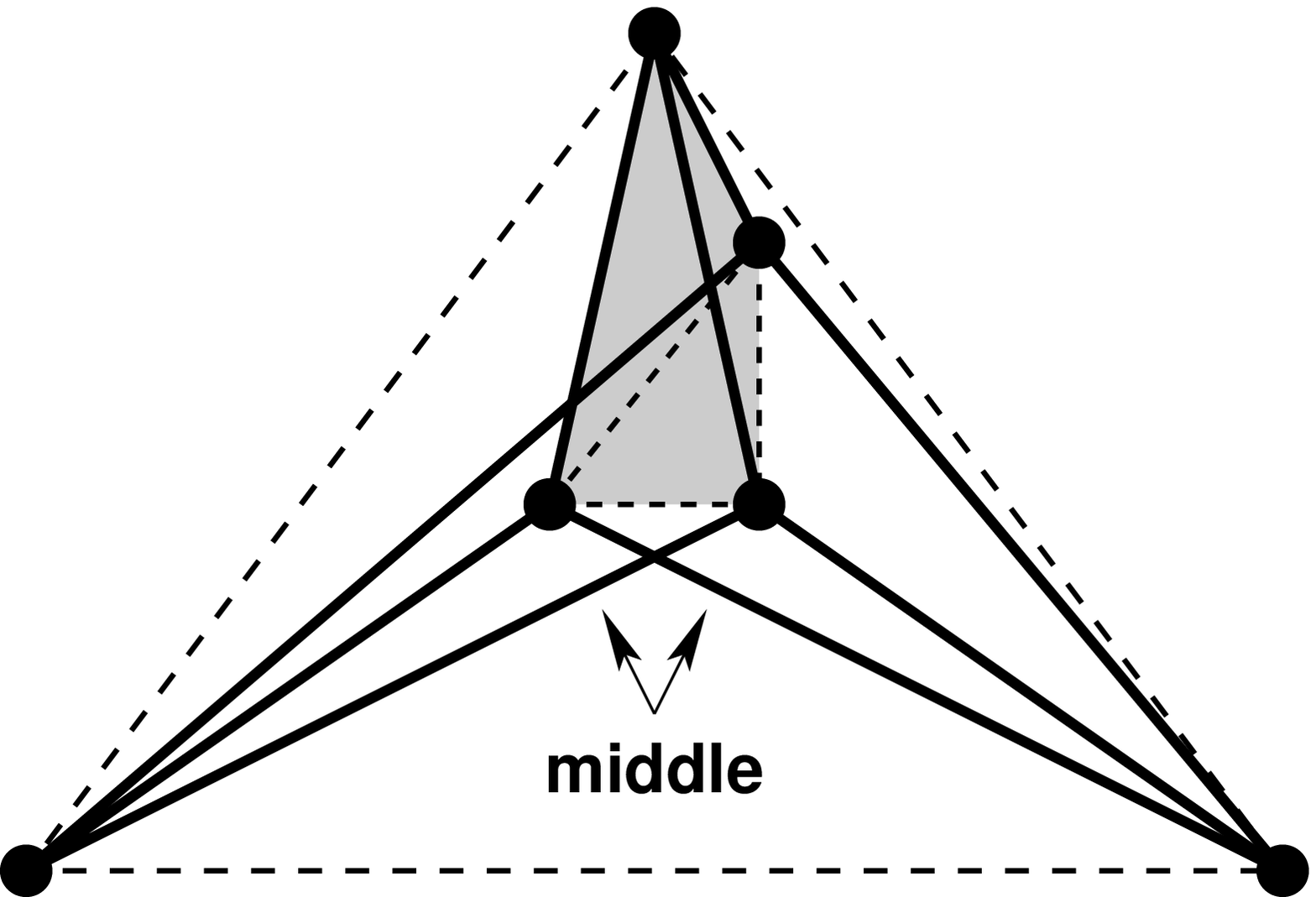}{0.25}}{figure}{fig:ccv_a}{Binary CCV}
The configuration CCV has two sub-configurations; the first, termed
\myem{unary}, has one middle vertex that is shared by both concave
kites; see Figure~\ref{fig:ccv_s}.  The second, termed \myem{binary},
has two distinct middle vertices belonging to each of the concave
kites; see Figure~\ref{fig:ccv_a}.

\begin{theorem}\label{thm:config}
A nested triangle drawing of $K_6$ belongs to one of the four
configurations: CCC, CCV, CVV or VVV.
\end{theorem}
\begin{proof}
According to~\cite{HoDCM} there are exactly two different rectilinear
drawings of $K_4$, of which the convex hull is either a triangle or a
quadrilateral.  The former has no crossings and corresponds to the
concave kite.  The latter has one crossing and corresponds to the
convex kite.

Since the drawing is comprised of nested triangles, a kite originates 
at each of the three outer vertices.  Since the vertices are non-collinear,
each of the kites is either convex or concave.  The drawing can have,
zero (CCC), one (CCV), two (CVV), or three (VVV) convex kites, with the
rest being concave.  
\end{proof}

\begin{lemma}\label{lem:cfg_concave}
If $m$ is a middle vertex of a concave kite in a nested triangle
drawing of $K_6$, then $m$ is contained within a quadrilateral 
composed of kite edges.
\end{lemma}
\begin{proof}
\figbox[l]{\boxeps{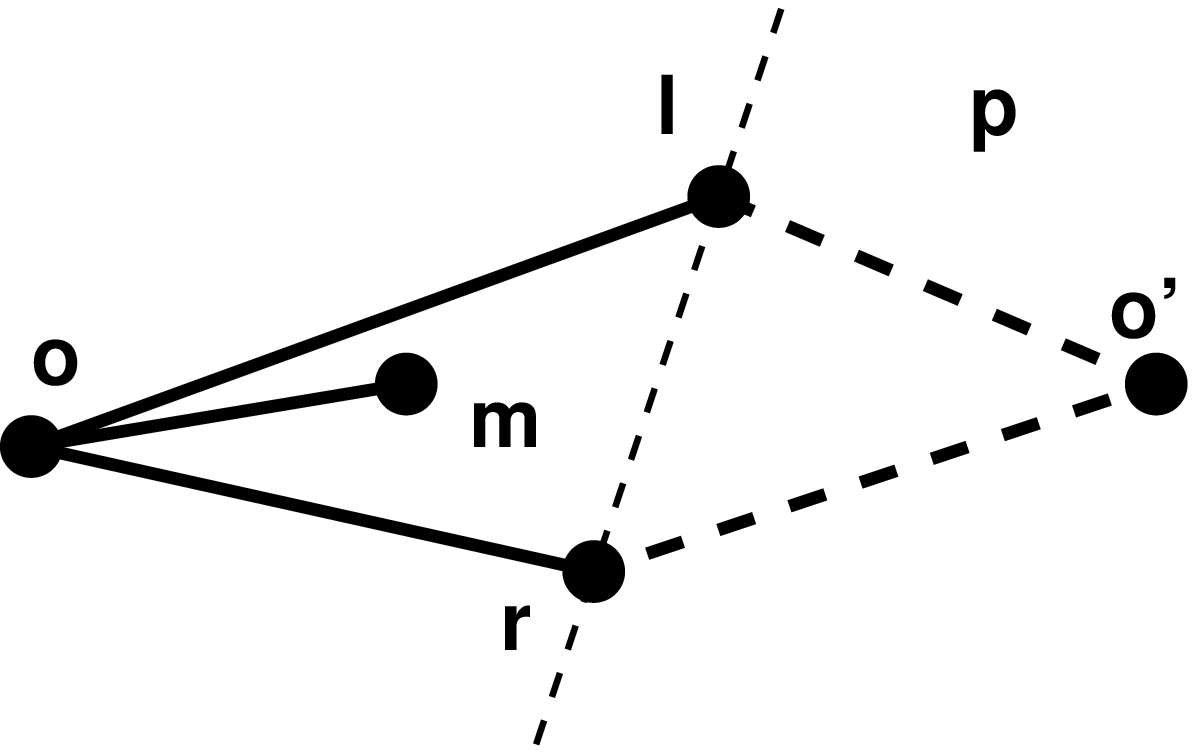}{0.25}}{figure}{fig:contain}{}
Let $\kappa$ be a concave kite in a nested triangle drawing with the
standard vertex labels $o$, $l$, $m$, and $r$.  Since $\kappa$ is
concave, the middle vertex $m$ is within the triangle $\Delta lor$.
The vertices $l$ and $r$ determine a line that defines a half-plane $p$ 
that does not contain $\kappa$.  Since the vertices $l$, $m$, and $r$
comprise the inner triangle of the drawing and must be contained within
the outer triangle, there must be an outer triangle vertex located
in the half-plane $p$.  Denote this vertex by $\op$ and note that a
kite originates from it; hence, there are kite edges $(\op,l)$ and
$(\op,r)$.  Thus, $m$ is contained within the quadrilateral
$(o,l,\op,r)$.  
\end{proof}

\begin{corollary}\label{cor:contain}
If $m$ is a middle vertex of a concave kite in a nested triangle 
drawing of $K_6$ and an edge $(v,m)$, originating outside the 
drawing, is incident on $m$, then the edge $(v,m)$ must cross one of
the kite edges.
\end{corollary}

\begin{remark}[Containment Argument]\label{rem:cont_arg}
Lemma~\ref{lem:cfg_concave} uses what will henceforth be referred to 
as the containment argument.  Consider two vertices contained in a
polygon.  These vertices define a line that bisects the plane.
In order for these vertices to be contained within the polygon, the two
half-planes must each contain at least one vertex of the polygon.
Similarly, if a vertex is contained inside two nested polygons and has
edges incident on all vertices of the outer polygon, then at least two
distinct edges of the inner polygon must be crossed by edges incident on
the contained vertex.
\end{remark}

\begin{lemma}[Barrier Lemma]\label{lem:barrier}
Let $o_1$, $o_2$, and $o_3$ be the outer vertices of a nested triangle
\figbox[r]{\boxeps{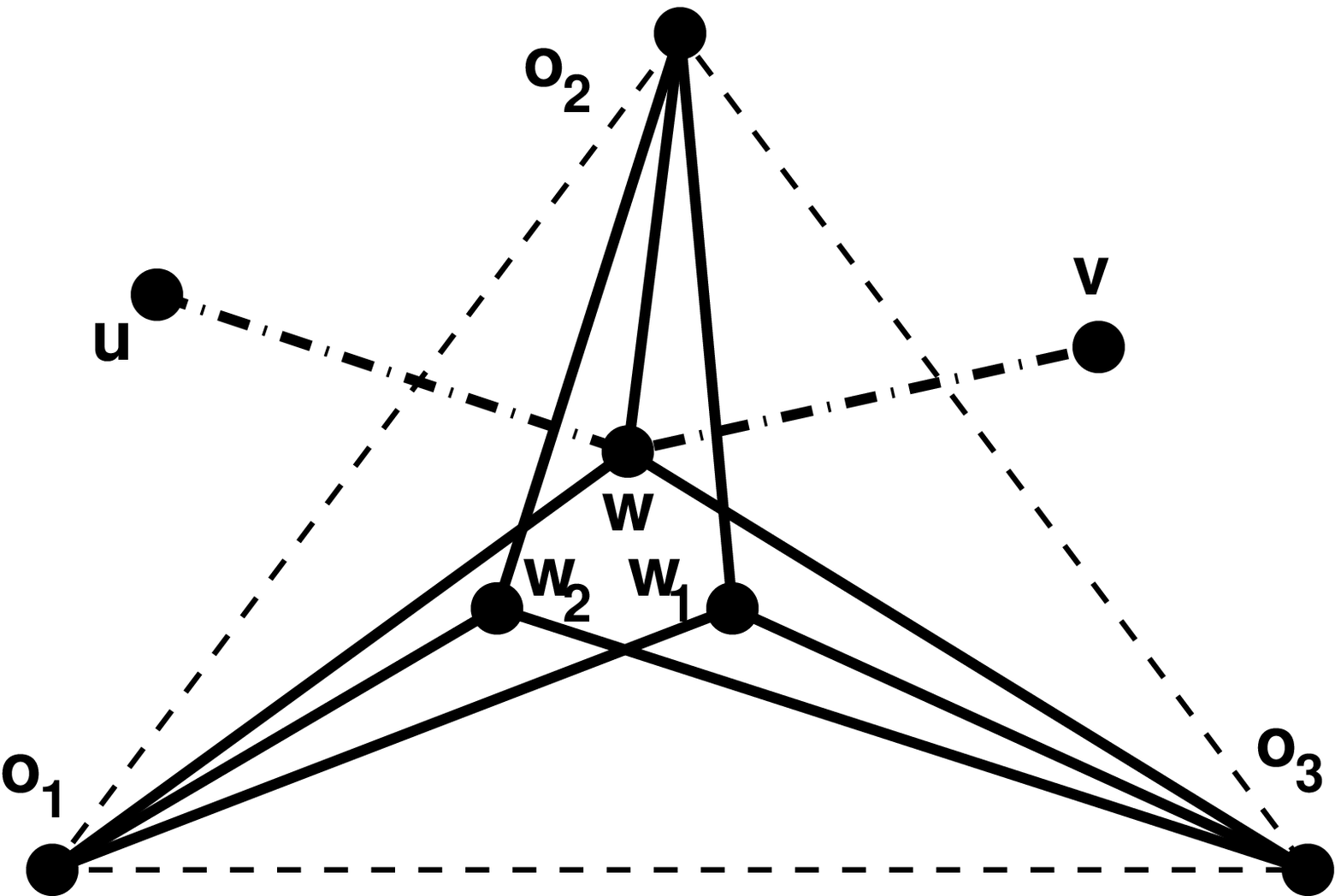}{0.25}}{figure}{fig:barx}{}
drawing of $K_6$, let $w$ be an inner vertex of the drawing, and
let $u$ and $v$ be two additional vertices located outside the outer 
triangle of the drawing.  If the edge $(u,w)$ crosses $(o_1,o_2)$
and the edge $(v,w)$ crosses $(o_2,o_3)$, then the total number of
kite edge crossings contributed by $(u,w)$ and $(v,w)$ is at least 
two.
\end{lemma}
\begin{proof}
\figbox[r]{\boxeps{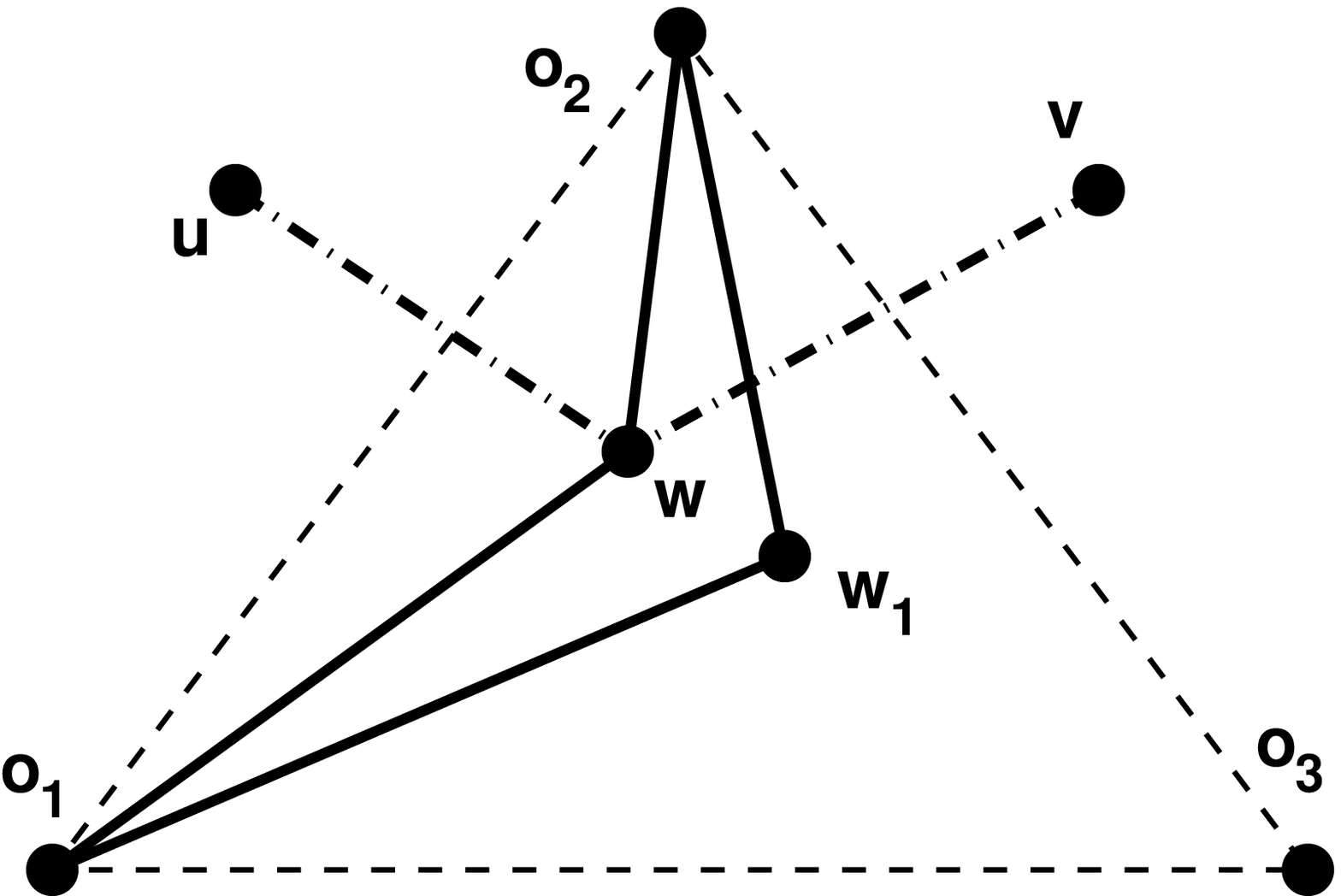}{0.25}}{figure}{fig:barrier}{}
If both edges $(u,w)$ and $(v,w)$ each cross at least one kite edge,
then we are done.  Without loss of generality, assume that $(u,w)$ does
not cross any kite edges.  Let $w_1$ and $w_2$ be the other two
inner vertices, and consider the path $(o_1,w_1,o_2)$.  Since edge
$(u,w)$ does not intersect the path, $(o_1,w_1,o_2)$ creates a barrier on 
the other side of path $(o_1,w,o_2)$.  The same argument with edge 
$(u,w)$ applies to path $(o_1w_2o_2)$, hence two barriers are present, 
forcing two crossings.
\end{proof}

To deal with the unary CCV configuration, see Figures~\ref{fig:ccv_s} and
\ref{fig:ccv_match}, we need to say something about the orientation of
the kites.  In a unary CCV configuration, the labels of the internal
vertices of the two concave kites must match; given a label, left,
middle, or right, and a vertex, it is impossible to distinguish one
concave kite from the other.  For example, the left vertex of one concave 
\figbox[l]{\boxeps{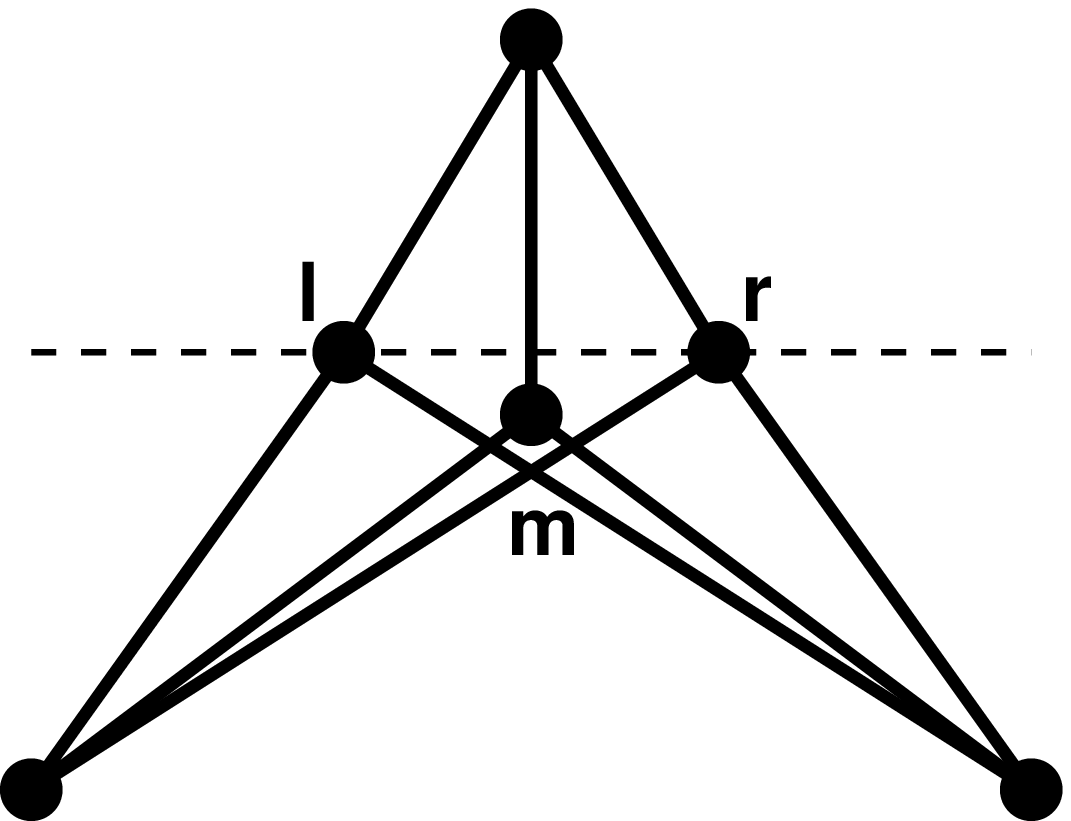}{0.25}}{figure}{fig:ccv_match}{Inside the
unary CCV}
kite is also the left vertex of the other concave kite.

\begin{lemma}\label{lem:shared}
If a nested triangle drawing of $K_6$ is in a unary CCV configuration,
then all three internal vertices of the two concave kites share the same 
labels.
\end{lemma}
\begin{proof}
Since the two concave kites share the same middle vertex, there are two
possible cases.  Either the labels of the internal vertices match, in 
which case we are done.  Otherwise, the left and right labels are 
interchanged.  By way of contradiction, assume that they are 
interchanged; this implies that the kites are disjoint, i.e. do not
overlap.  Consequently, they cannot share the middle vertex that is inside
both of the kites; this is contradiction.
\end{proof}

Lemma~\ref{lem:shared} implies that both concave kites are in the
half-plane defined by their left and right vertices, which contains the
shared middle vertex.  Moreover, by the containment argument 
(Remark~\ref{rem:cont_arg}), the convex kite must be in the other
half-plane.  Furthermore, no two kites in a CCC configuration share a
middle vertex.  

Just like the Barrier Lemma, the Kite Lemma, CCC Lemma, and $K_5$
Principle Lemma, are general lemmas that are used to derive properties
\figbox[l]{\boxeps{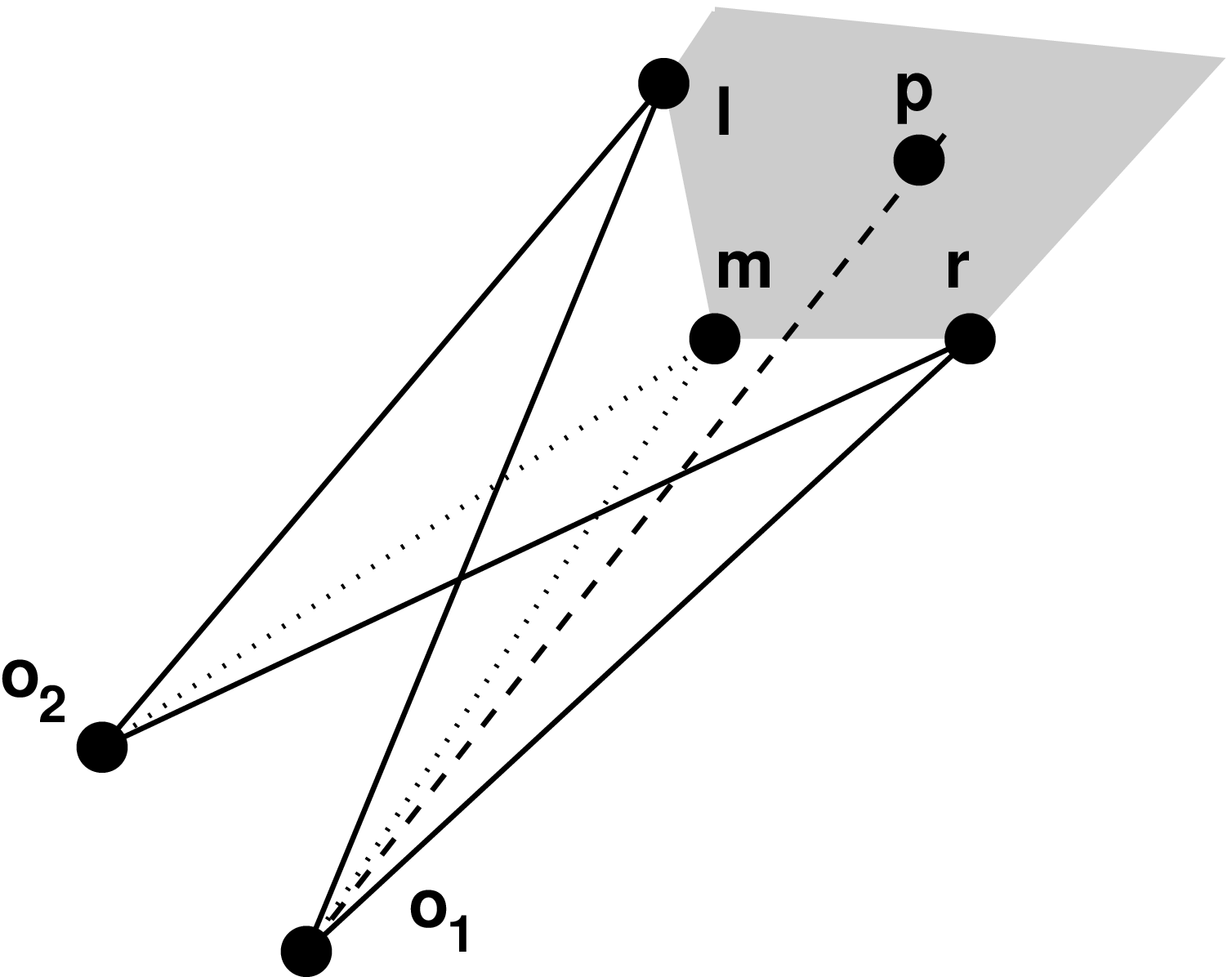}{0.25}}{figure}{fig:2kites}{}
of specific drawings.

\begin{lemma}[Kite Lemma]\label{lem:kites}
Let $\kappa_1 = (o_1,l,m,r)$ and $\kappa_2 = (o_2,l,m,r)$ be two concave 
kites such that they share the same internal vertices, the internal
vertices are labeled identically, and kite $\kappa_2$ does not contain 
vertex $o_1$ within it.  Let $A$ be the intersection of the sectors give 
by $<\!\!lo_1r$ and $<\!\!lmr$.  If $p$ is a vertex located in region $A$ 
and is noncollinear with any other pair vertices, then the edge $(o_1,p)$ 
must cross edge $(o_2,l)$ or edge $(o_2,r)$.
\end{lemma}  
\begin{proof}
Either vertex $o_2$ is contained in kite $\kappa_1$ or not.  If $o_2$ is
inside $\kappa_1$, then, because kite $\kappa_2$ is concave, a barrier 
path $(l,o_2,r)$ is created between vertex $o_1$ and vertex $p$.  Hence,
edge $(o_1,p)$ must cross the path $(l,o_2,r)$, intersecting one of the
path's two edges.

\figbox[r]{\boxeps{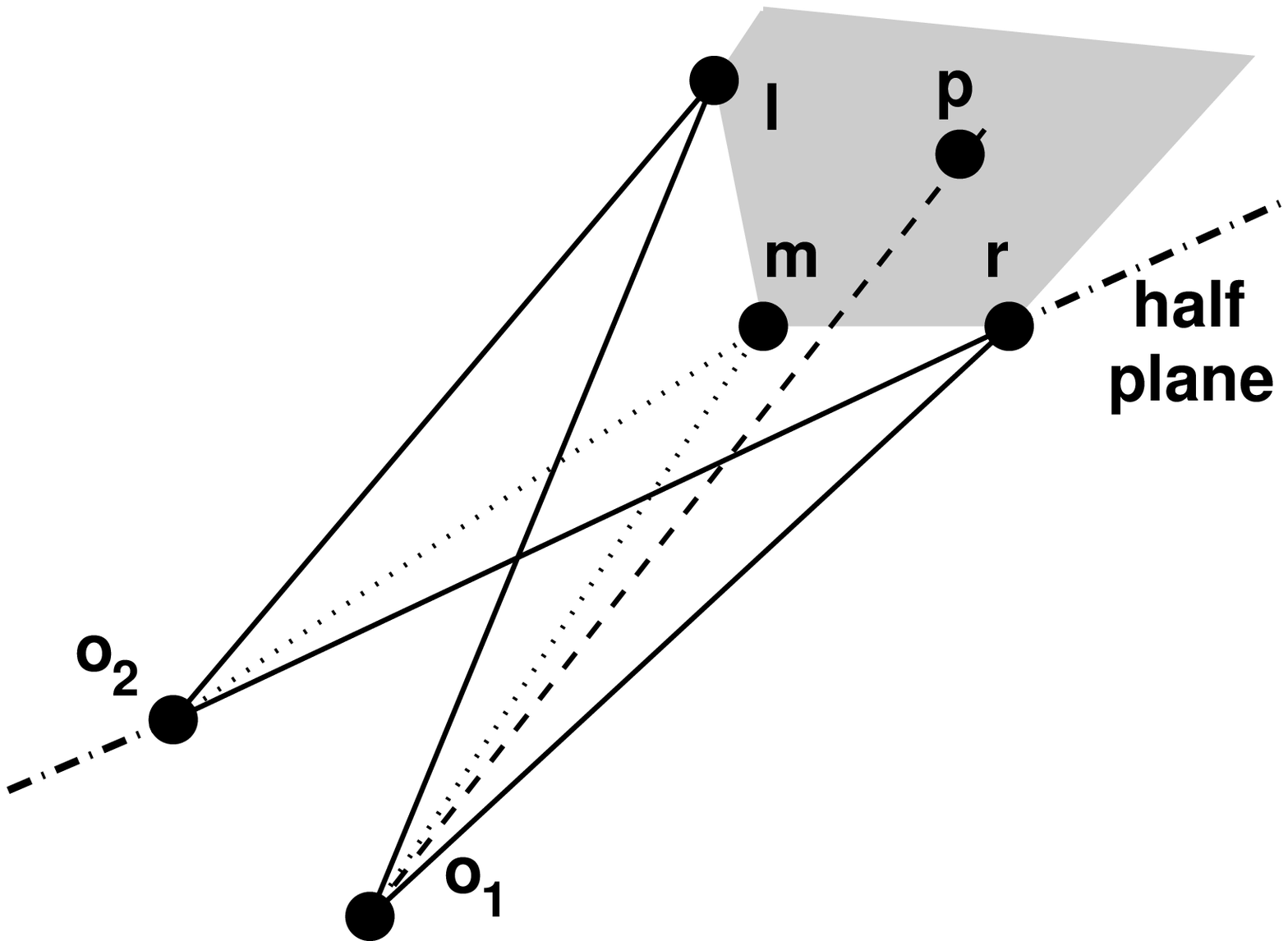}{0.25}}{figure}{fig:sidekite}{}
If vertex $o_2$ is not contained in kite $\kappa_1$, then assume, that
vertex $o_2$ is on the left side of kite $\kappa_1$ (clockwise with
respect to $o_1$).  The edge $(o_2,r)$ defines a half-plane that
separates vertex $p$ from vertex $o_1$.  Furthermore, the segment
defining the half-plane located within the sector $<\!\!lo_1r$
corresponds to part of the edge $(o_2,r)$.  Since the edge $(o_1,p)$
must be within the sector $<\!\!lo_1r$, it must cross edge $(o_2,r)$.

If vertex $o_2$ is on the right, by a similar argument, the edge
$(o_1,p)$ will cross edge $(o_2,l)$.
\end{proof}

\begin{lemma}[CCC Lemma]\label{lem:ccc_lem}
Given three kites in a CCC configuration,
denote the internal vertices $i_1$, $i_2$, $i_3$, and outer vertices
$o_1$, $o_2$, $o_3$ such that the middle vertex of a kite originating
\figbox[r]{\boxeps{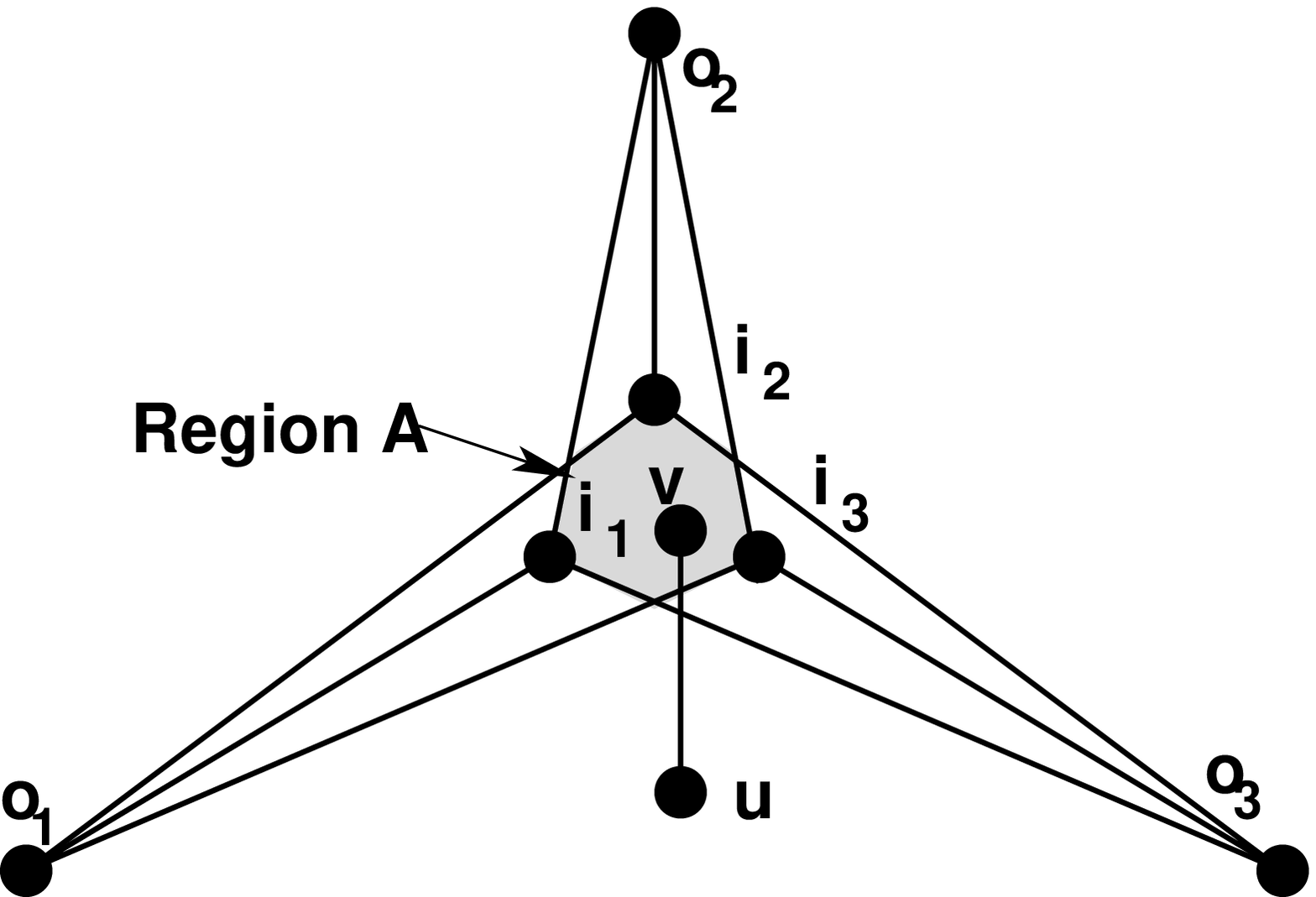}{0.25}}{figure}{fig:ccc_setup}{}
at $o_j$ is $i_j$.  Let $A$ be the region defined by the intersection
of sectors $<\!\!i_1o_2i_3$, $<\!\!i_2o_3i_1$, and $<\!\!i_3o_1i_2$.
Let vertex $u$ not be contained in any kite, let vertex $v$ be located in 
region $A$, and assume that no three vertices are collinear.  The edge 
$(u,v)$ must cross at least two kite edges.
\end{lemma}
\begin{proof}
\figbox[r]{\boxeps{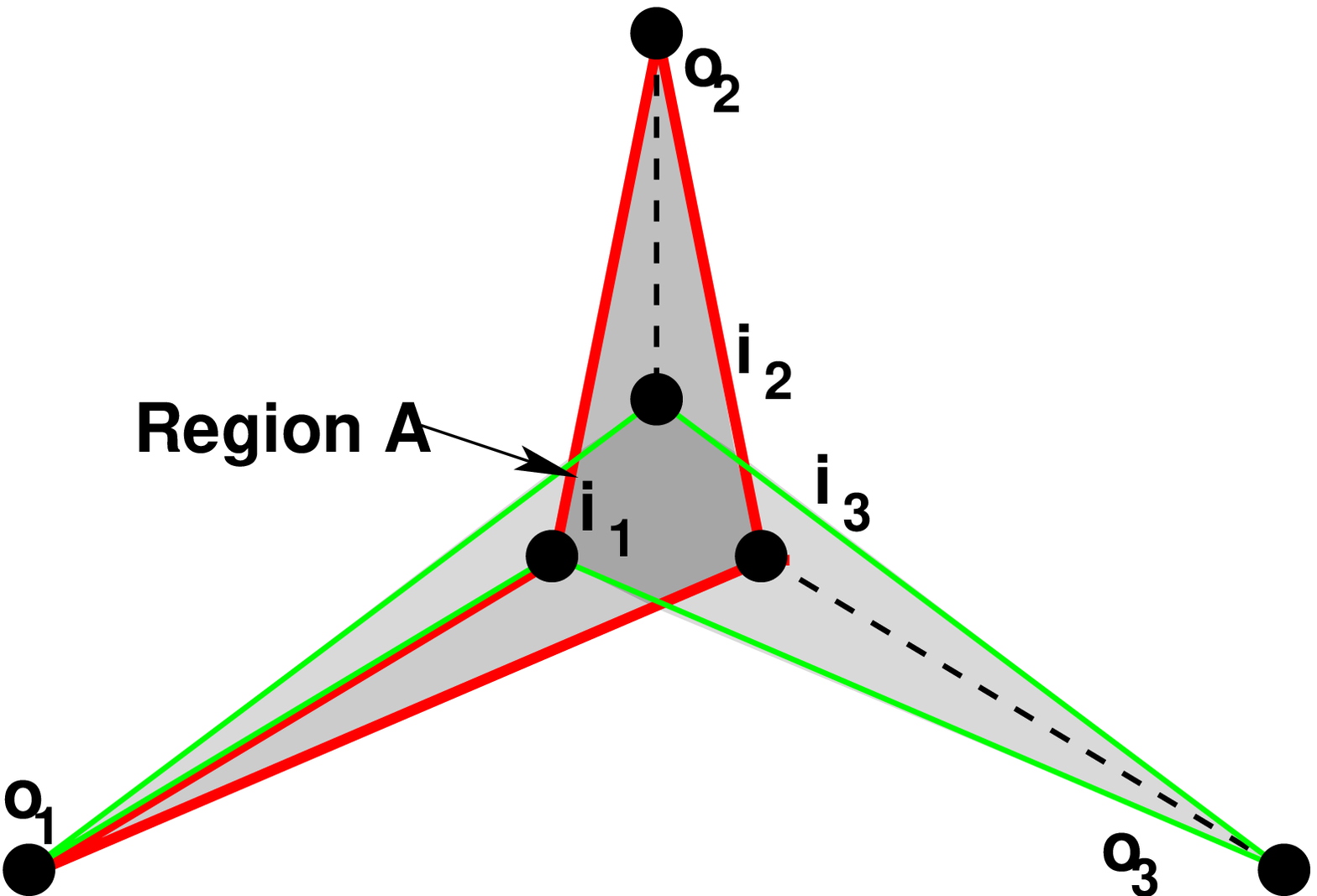}{0.25}}{figure}{fig:ccc_lem}{}
Using the kite edges we construct two polygons $(o_3,i_2,o_1,i_1,o_3)$
and $(o_2,i_3,o_1,i_1,o_2)$.  Since both polygons contain region $A$ and
since the only shared edge, is a middle edge, edge $(u,v)$ must cross
into both polygons, contributing at least one kite edge crossing from
each.  
\end{proof}

\begin{lemma}[$K_5$ Principle]\label{lem:k5_princ}
Let a drawing of $K_n$ have a triangular convex hull with the hull
coloured red and $n-3$ vertices contained within it coloured green.
The drawing has exactly $\binom{n-3}{2}$ \Xng{rg}{rg}\ edge crossings.
\end{lemma}
\begin{proof}
\figbox[l]{\boxeps{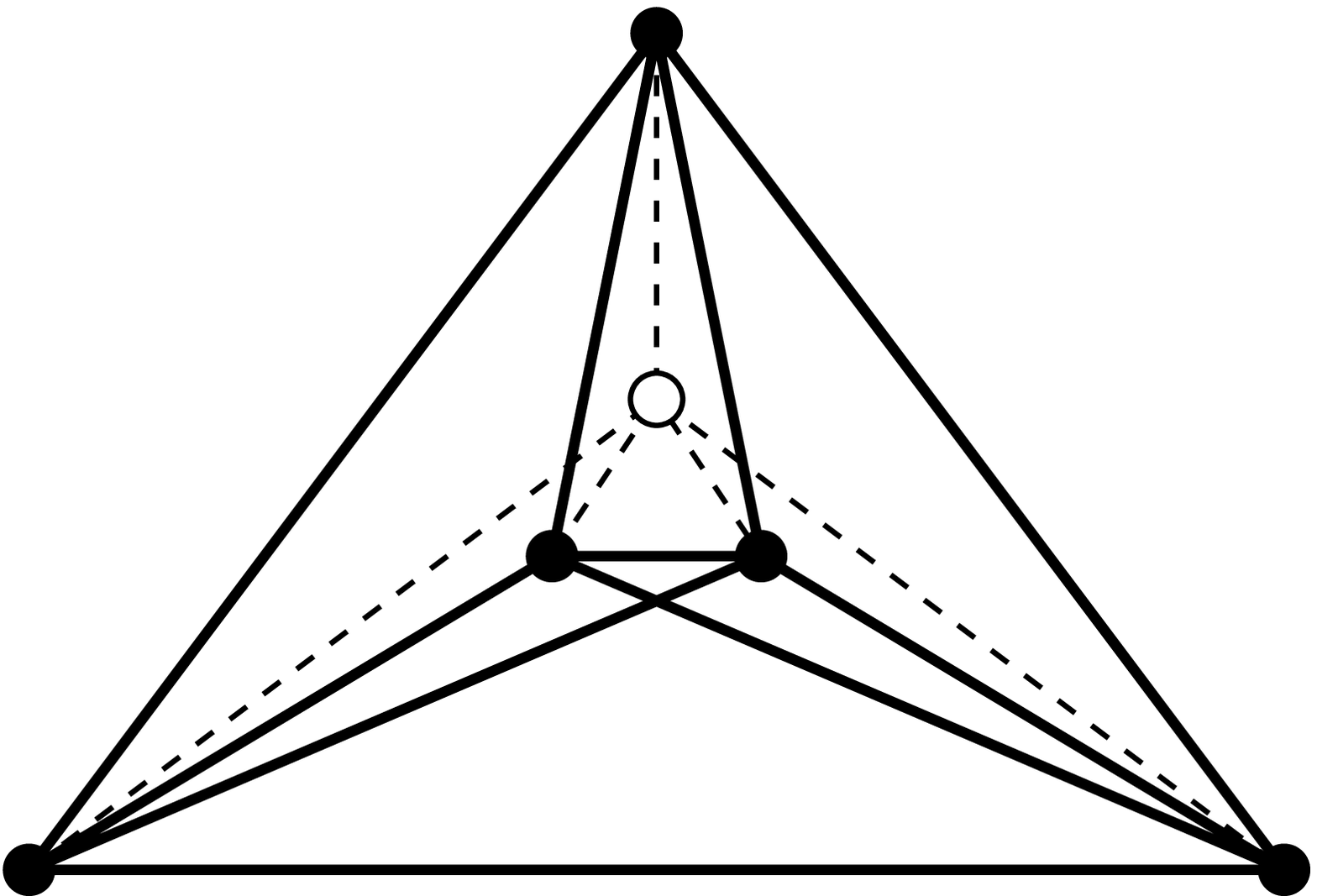}{0.25}}{figure}{fig:k5princ}{$K_5$ principle}
Select a pair of green vertices and remove all other green vertices
from the drawing.  This forms a $K_5$ with exactly one \Xng{rg}{rg}\
edge crossing that is uniquely identified by the two green vertices.
Since there are $\binom{n-3}{2}$ pairs of green vertices, there must be
$\binom{n-3}{2}$ \Xng{rg}{rg}\ edge crossings.  
\end{proof}

\section{The Proof}
Using configurations to abstract the vertex positions in drawings we
are now ready to combinatorially compute $\reccr{K_9}$ and
$\reccr{K_{10}}$.  We first reproduce the results from \cite{Si71} and
\cite{Gu72} proving that $\reccr{K_9} = 36$ and use these results to
show that $\reccr{K_{10}} = 62$.

The argument is as follows:  
\begin{enumerate}
\item Since $\reccr{K_{10}} \geq 61$, assume $\reccr{K_{10}} = 61$.
\item If $\reccr{K_{10}} = 61$ then the convex hull of an optimal of 
      $K_{10}$ must be a triangle.
\item If the convex hull of a drawing of $K_{10}$ is triangular then
      that drawing has 62 or more crossings, contradiction.
\item Therefore, $\reccr{K_{10}} \geq 62$
\end{enumerate}

\subsection{The Rectilinear Crossing Number of $K_9$}\label{sec:K_9}
We know from \cite{Si71} and \cite{Gu72} that the convex hull of an
optimal rectilinear drawing of $K_9$ must be a triangle.  By a
counting argument in~\cite{Si71}, the drawing must be composed of three
nested triangles, which we colour red, green, and blue.  Furthermore, the 
same paper argues that the red and green triangles are pairwise
concentric.  We derive these results for completeness.

As mentioned in the introduction, the rectilinear crossing numbers of
$K_6$ and $K_9$ are 3 and 36 respectively (see Table~\ref{tab:small_cr}); 
we make use of these facts throughout the following proofs.  We first
reproduce a result from~\cite{Si71} that states that an optimal
rectilinear drawing of $K_9$ must comprise of three nested triangles.

\begin{lemma}[Singer, \cite{Si71}]\label{lem:nested}
An optimal rectilinear drawing of $K_9$ consists of three nested 
triangles.
\end{lemma}
\begin{proof}
That the convex hull of an optimal rectilinear drawing of $K_9$ is a
triangle has been shown in~\cite{Gu72} and~\cite{Si71}.  Using a
counting technique similar to~\cite{Si71}, consider a drawing
composed of a red triangle that contains a green convex quadrilateral
that contains two blue vertices.  By the $K_5$ principle there are
$\binom{4}{2} = 6$ \Xng{rg}{rg}\ crossings.  At least two \Xng{rg}{gg}\
crossings are present because a convex quadrilateral cannot be
concentric with a triangle.  Selecting one green and one blue vertex at
a time and applying the $K_5$ principle yields, $4\cdot2 = 8$
\Xng{rb}{rg}\ crossings.  Six \Xng{rb}{gg}\ crossings are due to the
red-blue edges entering the green quadrilateral.  Applying the $K_5$
principle to the blue vertices yields one \Xng{rb}{rb}\ crossing.
There are $2 + 4 = 6$
\figbox[r]{\mbox{ \begin{tabular}{|l|l|}\hline
Crossing        & Count \\ \hline
\Xng{rg}{rg}    & 6 \\
\Xng{rg}{gg}    & 2 \\
\Xng{rb}{rg}    & 8 \\
\Xng{rb}{gg}    & 6 \\
\Xng{rb}{rb}    & 1 \\
\Xng{gg}{gg}    & 1 \\
\Xng{gb}{gg}    & 4 \\
\Xng{gb}{gb}    & 2 \\
\Xng{rb}{gb/gg} & 8 \\ \hline
Total           & 38 \\ \hline
\end{tabular}}}{table}{tab:nested}{Crossing contributions}
\Xng{gb}{gb}\ and \Xng{gb}{gg}\ crossings; the green quadrilateral is
initially partitioned into four parts by one \Xng{gg}{gg}\ crossing,
adding the first blue vertex creates two \Xng{gb}{gg}\ and adding the
second vertex creates two more \Xng{gb}{gg}\ crossings and two
\Xng{gb}{gb}\ crossings.  This totals 30 crossings.  An additional
eight \Xng{rb}{gb}\ and \Xng{rb}{gg}\ crossings occur inside the green
quadrilateral, four per blue vertex, totaling 38 crossings, which is
greater than the optimal 36.  By a similar argument any drawing whose
second hull is not a triangle will also be non-optimal; see 
Appendix~\ref{sec:k9_app}.
\end{proof}

Lemmas \ref{lem:2colour}, \ref{lem:non_conc}, \ref{lem:rb-gg}, 
\ref{lem:rb-rg}, and \ref{lem:internal}, count the number of different
crossings in an optimal drawing of $K_9$, making use of the nested
triangle property of Lemma~\ref{lem:nested}.

\begin{lemma}\label{lem:2colour}
A rectilinear drawing of $K_9$ comprising of nested triangles has a
minimum of three 2-coloured crossings of red-green, red-blue, and
green-blue.
\end{lemma}
\begin{proof}
Select two of the three red, green, and blue triangles.  These two
triangles form a nested triangle drawing of $K_6$ with three 2-colour
crossings.  Hence, there are three 2-colour edges of each type.
\end{proof}

\begin{lemma}\label{lem:non_conc}
A rectilinear drawing of $K_6$ comprising of nested non-concentric 
triangles has more than three crossings.
\end{lemma}
\begin{proof}
Let the outer triangle be red and the inner green.  By the $K_5$
Principle (Lemma~\ref{lem:k5_princ}) there are three \Xng{rg}{rg}\ edge
crossings.  If the two triangles are non-concentric then there is at
least one \Xng{rg}{gg}\ crossing.  
\end{proof}

\begin{lemma}\label{lem:rb-gg}
A rectilinear drawing of $K_9$ comprising of nested triangles has 
exactly nine \Xng{rb}{gg}\ crossings.  
\end{lemma}
\begin{proof}
The red triangle contains the green triangle and the green triangle
contains the blue triangle.  Therefore, every red-blue edge must cross
into the green triangle.  Since there are nine red-blue edges, there
are nine \Xng{rb}{gg}\ crossings.
\end{proof}

\begin{lemma}\label{lem:rb-rg}
A rectilinear drawing of $K_9$ comprising of nested triangles has at least 
nine \Xng{rb}{rg}\ crossings.  
\end{lemma}
\begin{proof}
The are three green and three blue vertices, thus there are nine unique 
green-blue pairs of vertices.  By the $K_5$ principle, each pair contributes
exactly one \Xng{rg}{rb}\ crossing.  Hence, a nested triangle drawing
of $K_9$ has exactly nine \Xng{rg}{rb}\ crossings.
\end{proof}

We call a crossing \myem{internal} if it is coloured either \Xng{rb}{gb}\ 
or \Xng{gb}{bb}.  The set of internal crossings consists of all
internal crossings in a drawing.  Intuitively, all internal
crossings take place within the green triangle.  We call a red-blue
kite \myem{full} if it contains a green vertex; otherwise we call it
\myem{empty}.  Intuitively a full red-blue kite contains a green-blue
kite.

\begin{lemma}\label{lem:internal}
The number of internal crossings in a nested triangle drawing of
$K_9$ is at least nine.
\end{lemma}
\begin{proof}
We make use of the fact that the green and blue triangles form a $K_6$
and that any rectilinear drawing of $K_6$ falls into one of the
five configurations: CCC, VVV, CVV, binary CCV, and unary CCV.  The
proof is by case analysis on the green-blue $K_6$ sub-drawing.  The
green-blue $K_6$ is drawn in one of the five configurations:

\noindent{\bf CCC configuration:\ }
Since each of the blue vertices is a middle vertex of a concave kite,
and all middle labels are distinct, by Corollary~\ref{cor:contain} each
of the nine red-blue edge crosses one green-blue edge, hence there are
nine \Xng{rb}{gb}\ crossings.

\noindent{\bf VVV or CVV configuration:\ }
If the drawing is in a VVV configuration, by the Barrier Lemma
(Lemma~\ref{lem:barrier}) there are two \Xng{rb}{gb}\ crossings per
blue vertex.  Adding the three \Xng{gb}{bb}\ crossings yields nine.  In
the CVV configuration one of the blue vertices is responsible for at
least three \Xng{rb}{gb}\ crossings rather than two; adding the two
\Xng{gb}{bb}\ crossings yields the required result.

\noindent{\bf Binary CCV configuration:\ }
Note that two of the blue vertices are responsible for three
\Xng{rb}{gb}\ crossings, and the third vertex is responsible for two.
\figbox[r]{\boxeps{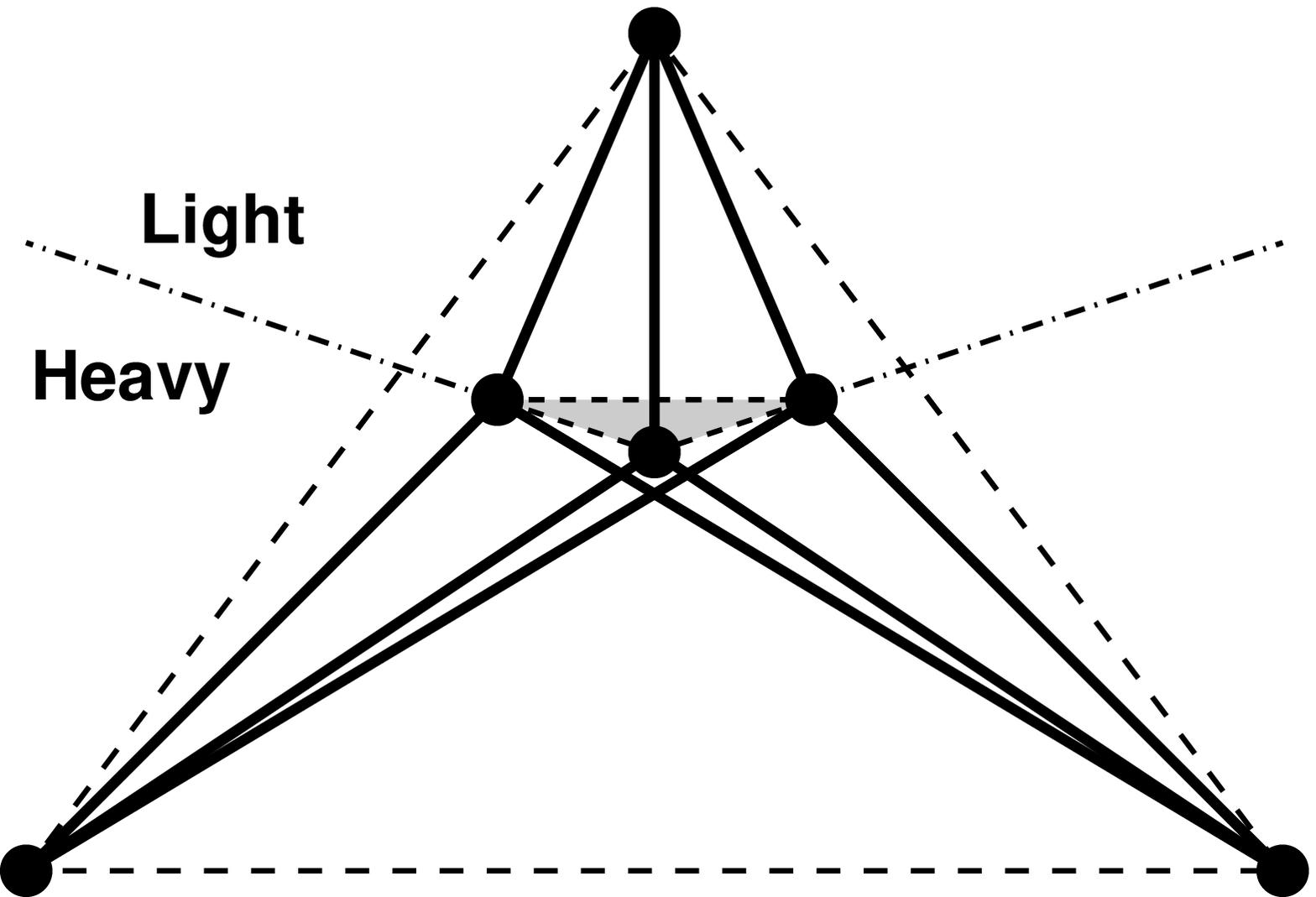}{0.25}}{figure}{fig:ccv_split}{Partition of
Drawing}
Adding the single \Xng{gb}{bb}\ crossing yields nine.

\noindent{\bf Unary CCV configuration:\ }
In the case of the unary CCV configuration, the drawing is
partitioned into a heavy and light part by extending the blue edges
incident on the middle vertex of the convex kite; see
Figure~\ref{fig:ccv_split}.  A red-blue kite whose origin vertex is in
the heavy side of the drawing is responsible for four or six
\Xng{rb}{gb}\ crossings while a red-blue kite originating in the light
side of the partition is responsible for three crossings if it is
empty, and one crossing if it is full; the six edge crossings occur if
there is an empty red-blue kite between the two concave kites.  In
order for the green triangle to be nested within the red, by the 
containment argument, at least one of the red-blue kites must originate 
in the heavy partition.  This implies that in order to get fewer than
\figbox[r]{\boxeps{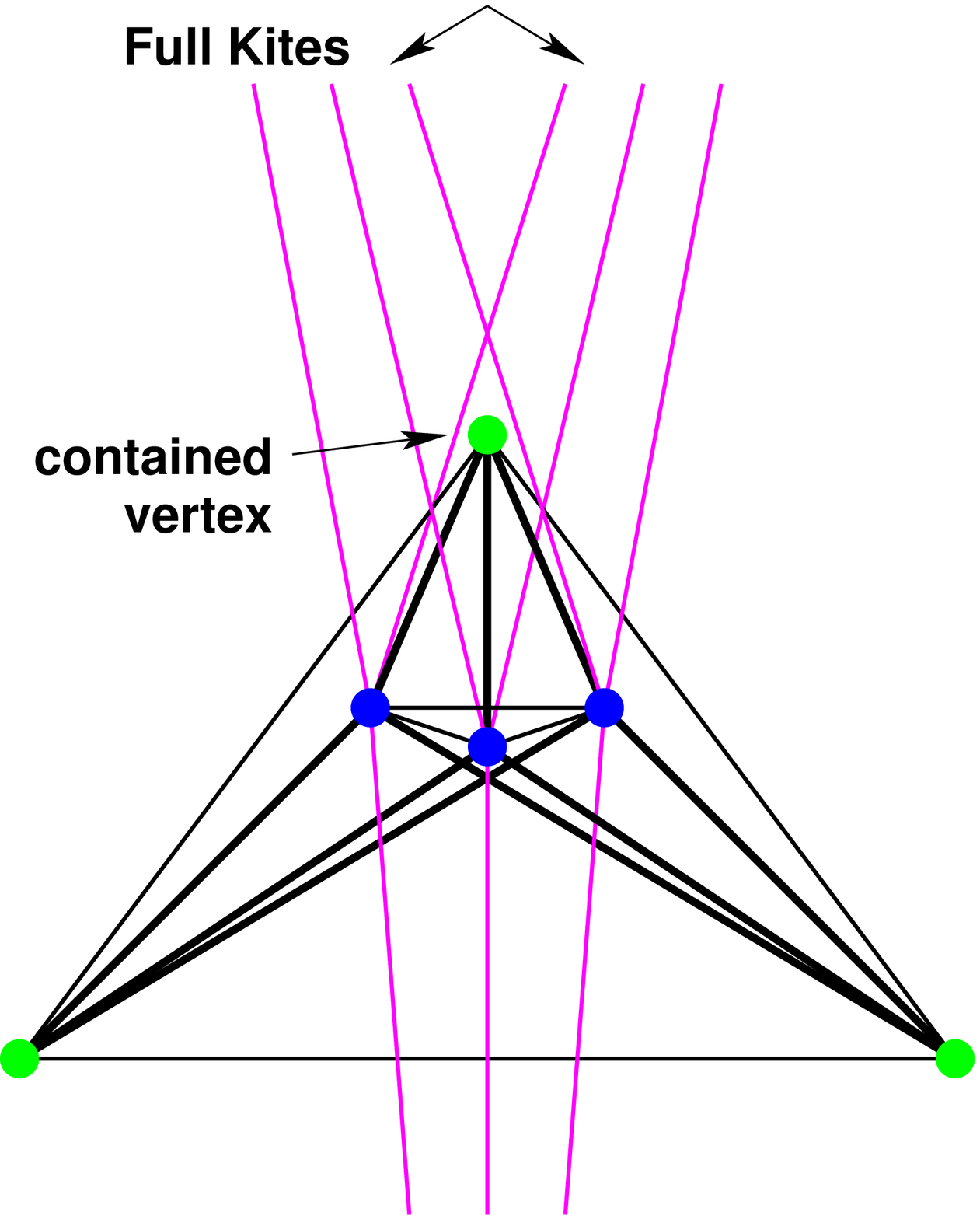}{0.25}}{figure}{fig:twofull}{}
eight \Xng{rb}{gb}\ crossings, two of the red-blue kites must be full and
contain the green-blue kite in the light partition.  This implies that
the third red-blue kite must be an empty kite between the two concave
red-green kites.  Since this kite is responsible for six crossings,
it follows that there are at least eight \Xng{rb}{gb}\ crossings and
therefore at least nine internal crossings.  \end{proof}

Singer's Theorem~\cite{Si71} follows from the previous lemmas.  A
stronger version of the theorem is given next.

\begin{theorem}\label{thm:optimal_k9}
An optimal rectilinear drawing of $K_9$ consists of three nested
triangles.  Furthermore, the red and green triangles, and the red and
blue triangles are concentric.
\end{theorem}
\begin{proof}
The first part of the statement is proven in~\cite{Gu72} and the
counting argument in Lemma~\ref{lem:nested}.

\figbox[l]{\mbox{ \begin{tabular}{|l|l|}\hline
Contribution                & Count \\ \hline
Lemma~\ref{lem:2colour}     & $\geq 9$ \\
Lemma~\ref{lem:rb-gg}       & $9$ \\
Lemma~\ref{lem:rb-rg}       & $9$ \\
Lemma~\ref{lem:internal}    & $\geq 9$ \\ \hline
Total                       & $\geq 36$ \\ \hline
\end{tabular}}}{table}{tab:optimal_k9}{Lower bound}
Putting Corollary~\ref{lem:2colour}, Lemma~\ref{lem:rb-gg}, and
Lemma~\ref{lem:rb-rg} together accounts for 27 of the 36 crossings in
an optimal drawing.  Lemma~\ref{lem:internal} states that there are
at least nine internal crossings.  Since $\reccr{K_9} = 36$, the number
of \Xng{rg}{gg}\ and \Xng{rb}{bb}\ crossings must be zero; this implies 
concentricity.
\end{proof}

\begin{corollary}\label{cor:nine_max}
An optimal rectilinear drawing of $K_9$ has at most nine \Xng{rb}{gb}\ 
crossings, at most two \Xng{gb}{bb}\ crossings and the total number of 
internal crossings is exactly nine.  
\end{corollary}
\begin{proof}
By Theorem~\ref{thm:optimal_k9}, an optimal drawing of a $K_9$ has 36
crossings.  Referring to Table~\ref{tab:optimal_k9}, an optimal
drawing has at least 27 non-internal edge crossings (Lemmas
\ref{lem:2colour}, \ref{lem:rb-gg}, and \ref{lem:rb-rg}).  By
Lemma~\ref{lem:internal}, there are at least nine internal edge
crossings and hence, an optimal drawing has exactly nine internal edge
crossings.

Three \Xng{gb}{bb} crossing occur if the green-blue $K_6$ part of the
drawing has configuration VVV.  However by a Barrier argument similar
to Lemma~\ref{lem:barrier} the configuration VVV creates nine
\Xng{rb}{gb}\ crossings plus three \Xng{gb}{bb}\ crossings, which
totals 12 internal crossings and cannot occur in an optimal drawing
of $K_9$.  Consequently at most two \Xng{gb}{bb}\ crossings may occur.
\end{proof}

\subsubsection{Optimal $K_9$ Drawings}
One is tempted to believe that an optimal drawing of $K_9$ is 
necessarily comprised of three nested triangles that are pairwise 
concentric.  However, this belief is fallacious, as is shown in
Figures~\ref{fig:k9_ccv} and~\ref{fig:k9_cvv}.
\figbox[h]{\boxeps{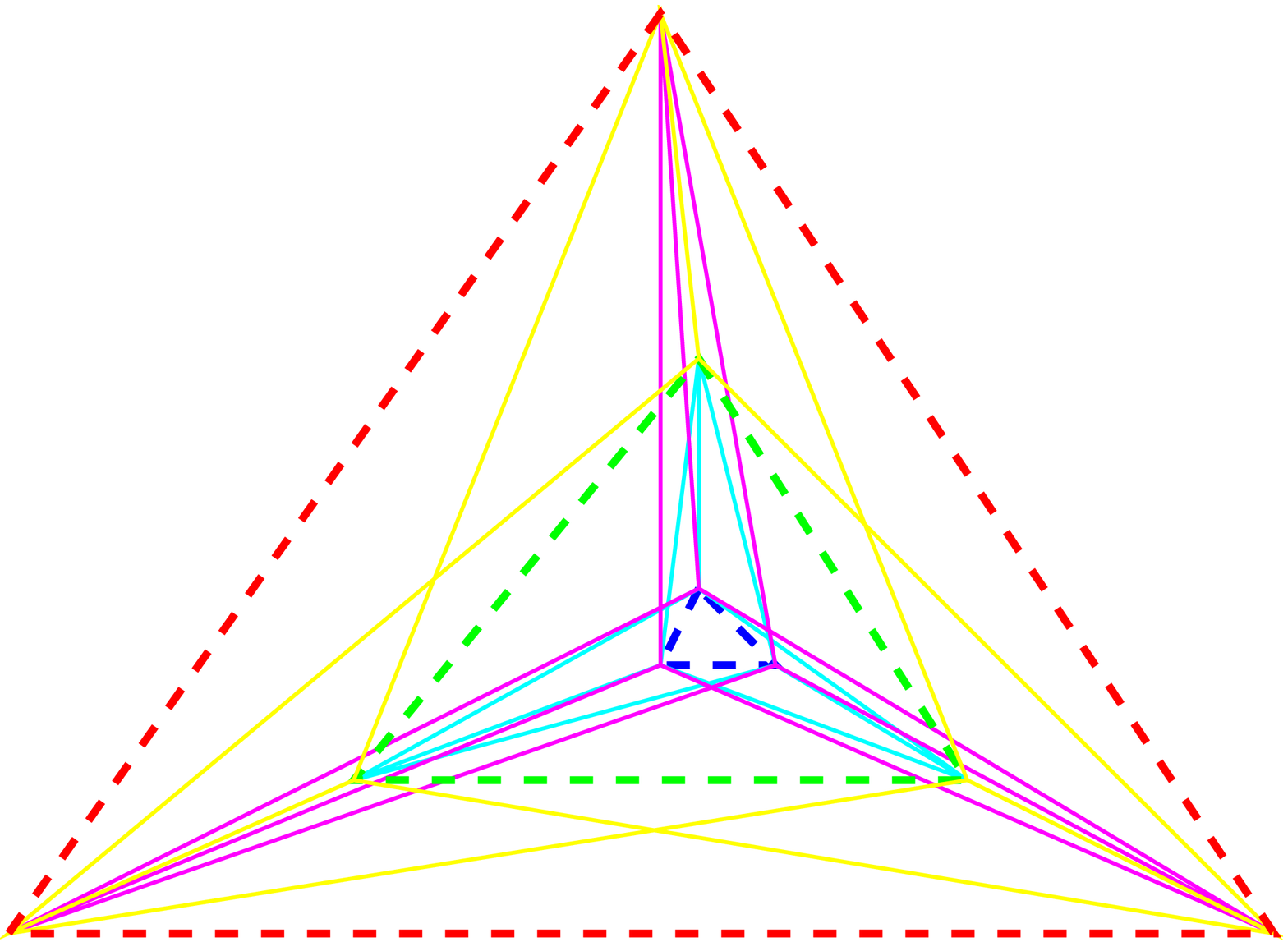}{0.35}}{figure}{fig:k9_ccc}{Blue-Green 
CCC Drawing}
\figbox[h]{\boxeps{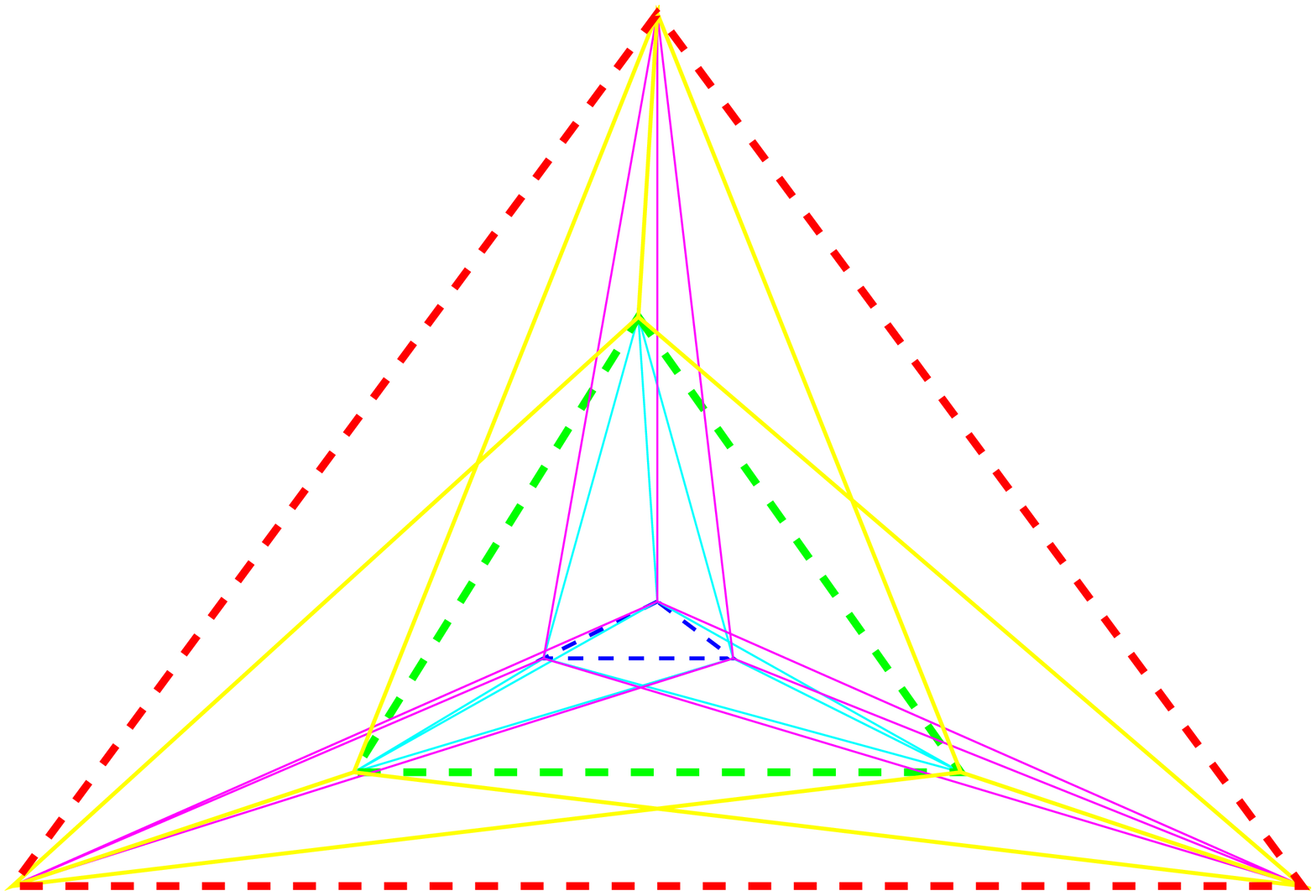}{0.35}}{figure}{fig:k9_ccv}{Blue-Green 
CCV Drawing}
\figbox[h]{\boxeps{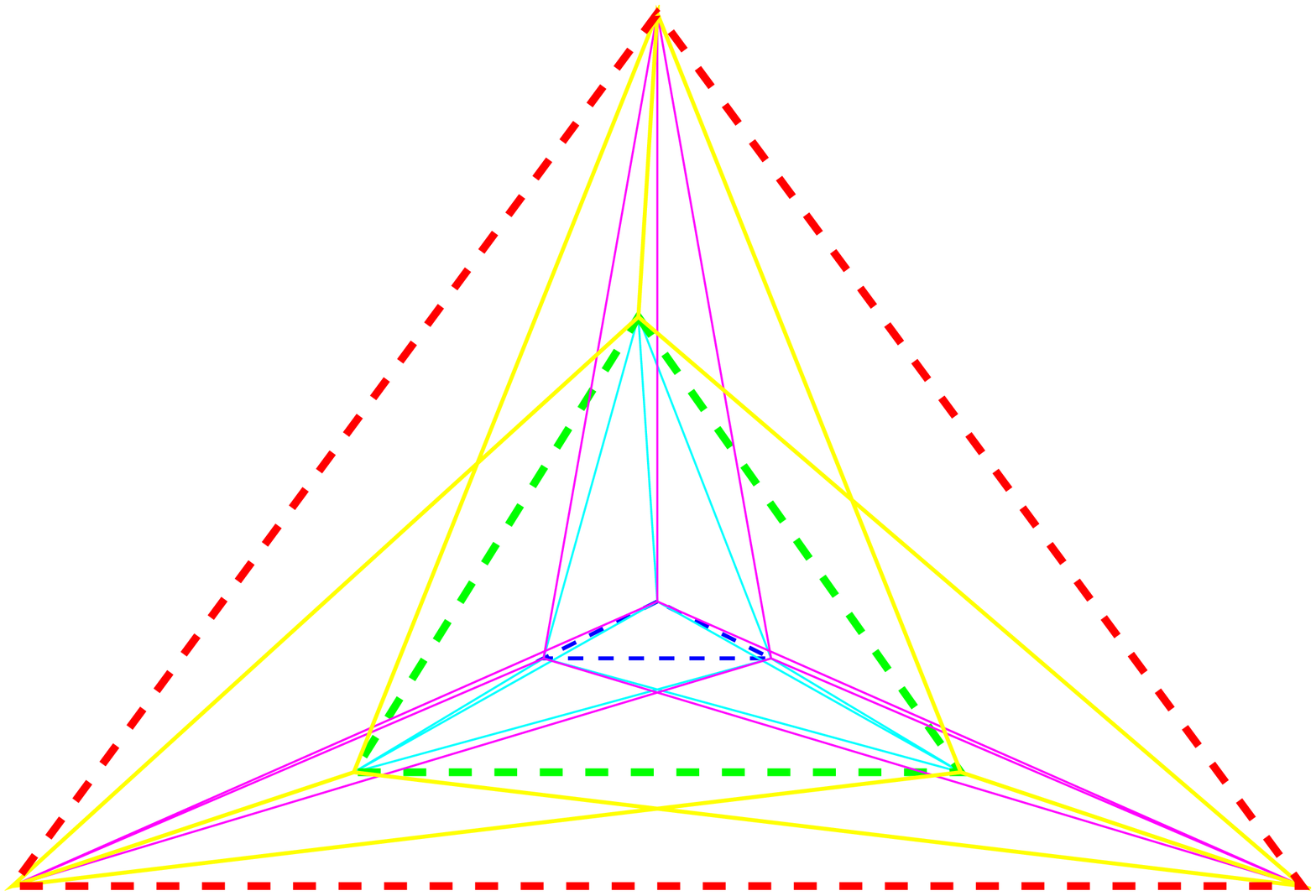}{0.35}}{figure}{fig:k9_cvv}{Blue-Green 
CVV Drawing}

\subsection{The Rectilinear Crossing Number of $K_{10}$}\label{sec:K_10}
We begin by reproducing a proof from~\cite{Si71} that $\reccr{K_{10}}
> 60$.  Since Singer~\cite{Si71,Ga86} exhibited a 62 crossing
rectilinear drawing of $K_{10}$, it follows that $61 \leq
\reccr{K_{10}} \leq 62$.

\begin{theorem}[Singer, \cite{Si71}]\label{thm:singer}
$\reccr{K_{10}} > 60$.
\end{theorem}
\begin{proof}
By way of contradiction, assume that there exists a rectilinear
drawing of $K_{10}$ with 60 crossings.  Since each edge crossing
comprises of four vertices, the sum of responsibilities of each vertex
totals $4 \cdot 60$.  Therefore, the average responsibility of each
vertex is $\frac{4 \cdot 60}{10} = 24$.  Furthermore, each vertex in
the drawing is responsible for exactly 24 edge crossings.  For if a
vertex is responsible for more than 24 edge crossings, then removing
the vertex from the drawing yields a drawing of $K_9$ with fewer than
36 edge crossings, which contradicts $\reccr{K_9}= 36$.  Similarly, if
the drawing has a vertex that is responsible for fewer than 24
crossings, then by the averaging argument, there must be a vertex that
is responsible for more than 24 crossings, leading to the same
contradiction.  Therefore, each vertex is responsible for 24
crossings.  Thus, any drawing of $K_{10}$ with 60 crossings contains an
optimal drawing of $K_9$.

Starting with an optimal drawing of $K_9$ we try to place the tenth 
vertex.  We have two choices; either place it such that one of the
hulls of the $K_{10}$ drawing is a convex quadrilateral or the
drawing comprises of nested triangles with a vertex in the inner
triangle.  In the latter case, the edge connecting the tenth vertex to
one of the outer triangle vertices must intersect an inner triangle
edge. Removing the inner triangle vertex that is opposite the
intersected edge creates a drawing of $K_9$ that fails the
concentricity condition.  Hence, the latter drawing will not be
optimal.  If the former situation arises there are two subcases.  If
the quadrilateral is the outer or the second hull, then removing an
inner vertex creates a non-optimal $K_9$ drawing, which is a
contradiction.  If the innermost hull is a convex quadrilateral, then
a priori it is not concentric with the outer triangle.  Let $b$ be
the vertex such that there is an edge from it to a vertex in the
outer triangle that intersects the quadrilateral.  Remove a vertex from
the quadrilateral that is antipodal to $b$.  This creates a non-optimal
$K_9$ drawing.  The result follows.  By an identical argument any 
rectilinear drawing of $K_{10}$ cannot have fewer than 60 crossings.
\end{proof}

Next, we study drawings of $K_{10}$ that have a nested triangle
sub-drawing of $K_9$ coloured in the standard way.  Let the tenth
vertex be coloured white; the responsibility of the tenth vertex is the
number of white crossings in the corresponding drawing of $K_{10}$.
The following technical Lemma is needed in the proof of Theorem
~\ref{thm:inner}.  This Lemma gives a lower bound on some of the white
crossings that occur within the green triangle.

\begin{lemma}\label{lem:rw-bw-x}
If a white vertex is added to a nested triangle drawing of $K_9$ such
that it is contained in the green triangle, then at least six crossings 
must exist of the types \Xng{rw}{gb}, \Xng{rb}{bw}, \Xng{gb}{bw}, and 
\Xng{rg}{gg}.
\end{lemma}
\begin{proof}
At least two of the red-white edges must cross into the green triangle
on distinct green-green edges as a consequence of the nested triangle
requirement and the containment argument.  Select two of the three
red-white edges such that they cross into the green triangle on distinct
green-green edges and such that the total number of \Xng{rw}{gb}\
crossings is minimized.  Let $c_1$ and $c_2$ be the number of
\Xng{rw}{gb}\ crossings for which each of the two red-white edges is
responsible and assume, without loss of generality, that $c_1 \leq c_2$.
The lower bound on the total number of \Xng{rw}{gb}\ crossings is $2c_1
+ c_2$.  We say that the red-white edge of lesser responsibility ($c_1$)
has \myem{weight two}, and we say that the other red-white edge, of
responsibility $c_2$, has \myem{weight one}.

Upon examining \Xng{rw}{gb}\ crossings the proof falls into three main 
cases corresponding to the numbers of \Xng{rw}{gb}\ crossings; if there
are six or more \Xng{rw}{gb}\ crossings then we are done.  We consider
the cases when the number of \Xng{rw}{gb}\ crossings is $\{0,1,2\}$, 
$\{3\}$, and $\{4,5\}$, the latter being the most challenging.

\noindent{\bf Case 1: 0, 1, or 2 \Xng{rw}{gb}\ crossings}\\
By the Barrier Lemma, every blue vertex forces at least one
\Xng{rw}{gb}\ crossing.  Hence, there must be at least three
\Xng{rw}{gb}\ crossings.

\noindent{\bf Case 2: 3 \Xng{rw}{gb}\ crossings}\\ 
Considering only the \Xng{rw}{gb}\ crossings, the configuration that
minimizes the number of \Xng{rw}{gb}\ crossings occurs when the
\figbox[r]{\boxeps{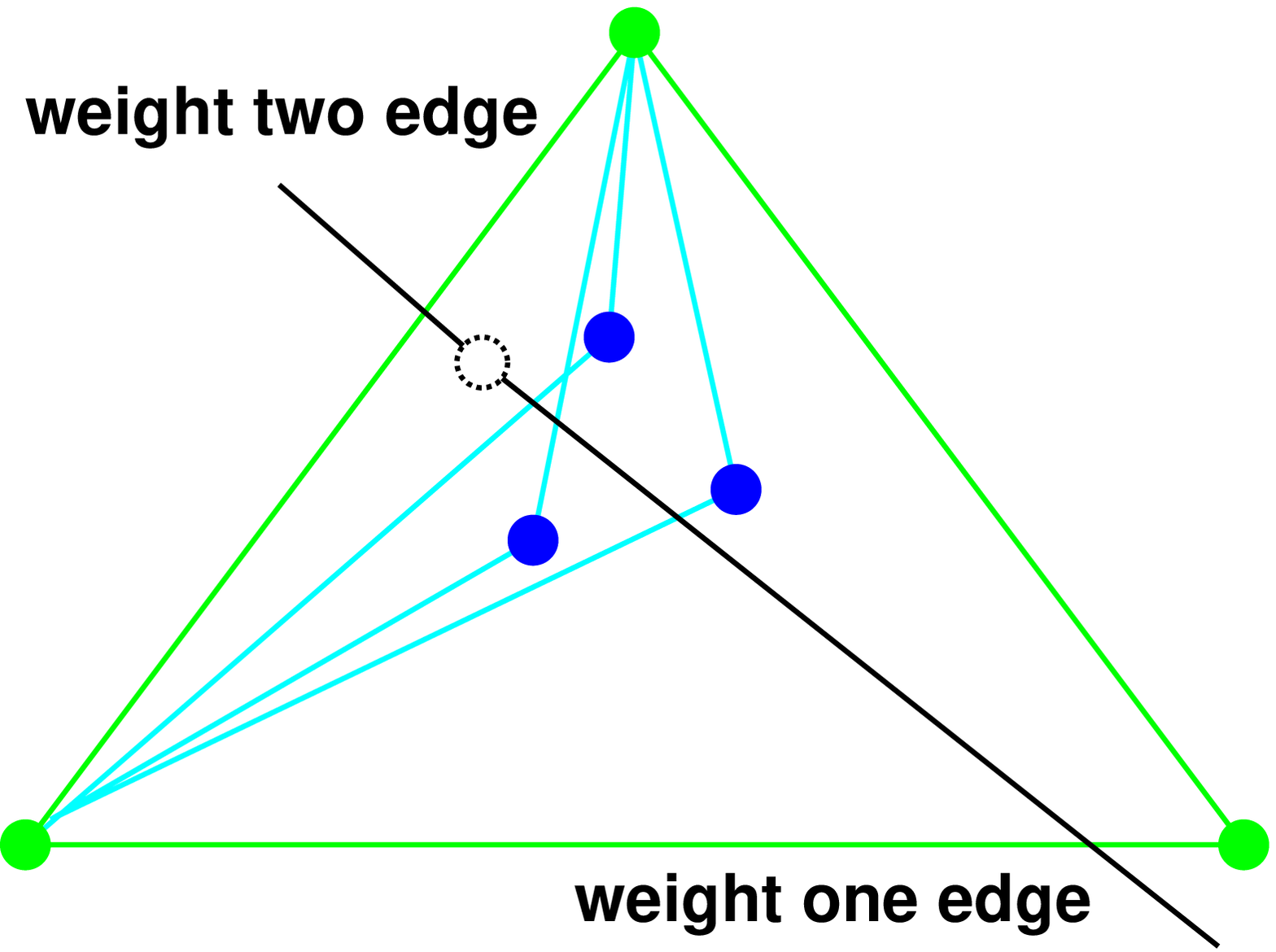}{0.25}}{figure}{fig:min3}{}
red-white edge of weight two crosses zero green-blue edges and the
red-white edge of weight one crosses three.  However, we must consider
blue-white edges also; by the Barrier principle one of the blue-white
edges must cross at least two green-blue edges, and the other must
cross at least one.  This brings the total up to at least six.

\noindent{\bf Case 3: 4 or 5 \Xng{rw}{gb}\ crossings}\\
Assume there are at least four \Xng{rw}{gb}\ crossings.  If there are
two or more \Xng{gb}{bw}\ crossings then we are done.  It remains to
consider two subcases: that of zero or one \Xng{gb}{bw}\ crossings.

\noindent{\bf Subcase 3.1: $0$ \Xng{gb}{bw}\ crossings} \\
Assume there are zero \Xng{gb}{bw}\ crossings.  This case can only occur 
when no green-blue edge intersects the blue triangle, i.e., the
\figbox[l]{\boxeps{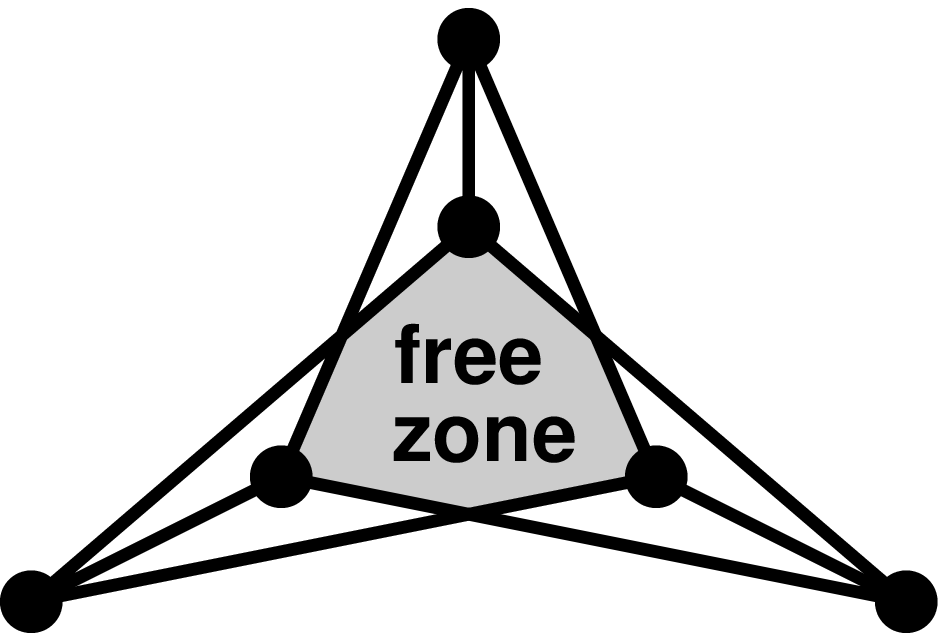}{0.25}}{figure}{fig:free}{}
green-blue kites are in a CCC configuration because there are no
\Xng{gb}{bb}\ crossings.  The white vertex is in the green-blue
free zone; a \myem{free zone} consists of all regions of a nested
triangle drawing of $K_6$ where a seventh vertex can be placed such
that no kite edge blocks visibility of any inner vertices.  Note that
removal of the inner edges of all convex kites in a configuration
creates a free zone.  A free zone occurs naturally in a CCC
configuration.   

If there is a green-blue edge intersecting the blue triangle, then
there exists a green-blue-green path between two of the blue vertices
that forces at least one \Xng{gb}{bw}\ crossing.  Since the white
\figbox[l]{\boxeps{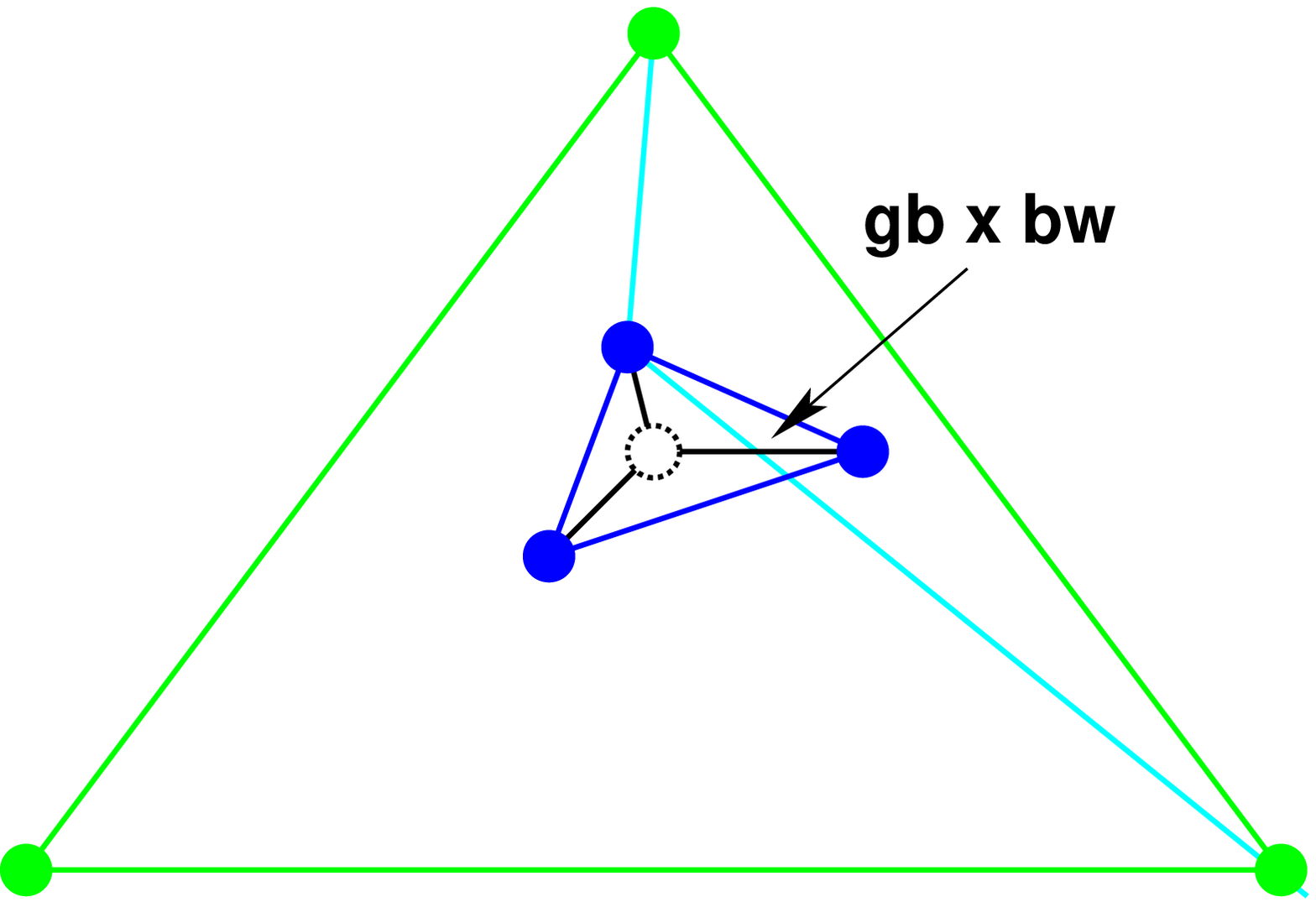}{0.25}}{figure}{fig:gbpart}{The path}
vertex must be in the naturally occurring green-blue free zone, i.e., a
green-blue CCC configuration, by the CCC Lemma (Lemma~\ref{lem:ccc_lem},
this forces every red-white edge to generate at least two \Xng{rw}{gb}\ 
crossings.  This yields a total of at least six crossings.

\begin{remark}
We reach a count of five crossings of the required type.  The remainder
of the proof is devoted to producing one more edge crossing of one of
the required types. 
\end{remark}

\noindent{\bf Subcase 3.2: $1$ \Xng{gb}{bw}\ crossing}\\
We now consider the \Xng{rb}{bw}\ crossings.  Consider the red-blue
kite configuration.  Either the configuration is a CCC or not.

\noindent{\bf Subcase 3.2.1: Non-CCC red-blue configuration}\\
Assume that the red-blue kite configuration is not in a CCC
configuration.  By the converse of the argument used in Subcase 3.1
there is at least one \Xng{rb}{bw}\ crossing.  Adding to the existing
five yields at least six distinct crossings of the required type.
This leaves only one case: the CCC red-blue configuration.

\noindent{\bf Subcase 3.2.2: CCC red-blue configuration}\\
We now consider the five subcases corresponding to the distinct
green-blue configurations within the red-blue CCC configuration.

\noindent{\bf Subcase 3.2.2.1: CCC green-blue configuration}\\
If the green-blue kites are in a CCC configuration, then this case is
covered by subcase 3.1.

\noindent{\bf Subcase 3.2.2.2: CVV and VVV green-blue configurations}\\
For every green-blue edge that intersects the blue triangle, there is
at least one \Xng{gb}{bw}\ edge crossing; see Figure~\ref{fig:gbpart}.  
Hence, if the green-blue kites are in a CVV or a VVV configuration then
we have at least two \Xng{gb}{bw}\ crossings.  This sums to at least
six crossings.

\noindent{\bf Subcase 3.2.2.3: Unary CCV green-blue configuration} \\
If the green-blue configuration is a unary CCV configuration then the red
and green triangles are not concentric; therefore, there is at least one
\Xng{rg}{gg}\ crossing.  Adding at least four \Xng{rb}{bw}\ crossings, and
at least one \Xng{gb}{bw} crossing, by the same argument as in subcase 
3.2.2.2, yields at least six crossings.

\noindent{\bf Subcase 3.2.2.4: Binary CCV green-blue configurations}\\
We are now left with the case of a CCC red-blue kite configuration and
a binary CCV green-blue kite configuration with the white vertex either
inside the red-blue free zone or not.

If the white vertex is not inside the red-blue free zone then there is
at least one \Xng{rb}{bw}\ crossing, by the same argument used in
subcase 3.1, plus at least one \Xng{gb}{bw} crossing, by the same
argument as in subcase 3.2.2.2, plus at least four \Xng{rw}{gb}\ 
crossings.  The sum of these crossings is at least six.  

Thus, assume that the white vertex is in the red-blue free zone.  We
will argue that there must always be either at least five
\Xng{rw}{gb}\ crossings plus at least one \Xng{gb}{bw}\ crossing, or at
least four \Xng{rw}{gb}\ crossings plus at least two
\Xng{gb}{bw}\ crossings.

Consider the drawing minus the single green-blue edge in the only
convex green-blue kite, i.e., the inner edge of the convex kite.  This
creates a green-blue free zone, inside of which there are no
\Xng{gb}{bw}\ edge crossings.

\begin{remark}\label{rem:usekites}
In order to cross into the green-blue free zone, a red-white edge must
cross a green-blue edge.  Furthermore, if a green-blue kite and a
red-blue kite are both concave, and have their internal (blue) vertices
labeled identically, then we may invoke Lemma~\ref{lem:kites} (Kite
Lemma).  That is, the red-white edge, incident on the origin vertex
(red) of the red-blue kite, must cross into the concave green-blue kite
before crossing into the free zone.  This produces an additional
\Xng{rw}{gb}\ crossing.  
\end{remark}

The white vertex is either inside the green-blue free zone or not.

If the white vertex is inside the green-blue free zone, then the
red-blue CCC configuration together with the pigeon-hole principle
implies that we can match up a concave red-blue kite with each of the
two concave blue-green kites.  By remark~\ref{rem:usekites}, each of
these match-ups contribute at least two \Xng{rw}{gb}\ crossings, and
the third red-white edge contributes at least one \Xng{rw}{gb}\ 
crossing.  Thus, if the white vertex is in the green-blue free zone
there are five \Xng{rw}{gb}\ crossings.  By the argument used in
subcase 3.2.2.2, the single convex green-blue kite contributes to at
least one \Xng{gb}{bw}\ crossing.  Thus we get at least six crossings.

If the white vertex is outside the green-blue free zone, then we get at
least one \Xng{gb}{bw}\ crossing by the same argument used in subcase 3.1
and at least one \Xng{gb}{bw}\ crossing by the same argument used in
subcase 3.2.2.2.  Since we have at least four \Xng{rw}{gb}\ crossings
(case 3), we get a grand total of at least six crossings.

In all possible cases that can occur we have shown that the number of
crossings of the required type is at least six.
\end{proof}

Lemma~\ref{lem:nested} imposed a nested triangle requirement on any
optimal rectilinear drawing of $K_9$.  The following lemma imposes a
similar constraint on optimal rectilinear drawings of $K_{10}$.

\begin{lemma}\label{lem:tri_hull}
If $\reccr{K_{10}} = 61$ then the first two hulls of an optimal 
rectilinear drawing of $K_{10}$ must be triangles.
\end{lemma}
\begin{proof}
By way of contradiction, assume that there exits an optimal rectilinear
drawing of $K_{10}$ whose convex hull is not a triangle and 61 edge
crossings.  By the same averaging argument used in
Theorem~\ref{thm:singer}, at least four of the vertices are responsible
for 25 edge crossings; removing any of them yields an optimal drawing
of $K_9$ with 36 crossings.  If any of the vertices with responsibility
25 are not on the convex hull, then removing such a vertex yields a
drawing of $K_9$ with a non-triangular convex hull, which is a
contradiction.  Therefore, all the vertices of responsibility 25 must
be on the convex hull of the original drawing.  Since we can always
remove one of the four vertices such that the outer hull of the new
drawing is not a triangle, this contradicts the original assumption.
Hence, the first convex hull must be a triangle.

Assume that the second hull is not a triangle.  Either the second hull
is a convex quadrilateral or the second hull has more than four
vertices; assume the latter.  Since at least four of the vertices must
have responsibility 25 and the outer hull is a triangle, at least one
vertex of responsibility 25 must either belong to the second hull, or
be contained within it.  In either case, removing said vertex creates
a drawing of $K_9$ that has 36 crossings and whose second hull is
not a triangle.  This is a contradiction.

Finally, assume that the second hull is a convex quadrilateral.
If within the second hull there is a vertex of responsibility 24 or
higher, removing said vertex creates a drawing of $K_9$ with 37 or
fewer vertices.  By Lemma~\ref{lem:nested} such a drawing should have
at least 38 crossings, contradiction.  Hence, assume that all three
vertices inside the second hull have responsibility 23.  Consequently,
the remaining 7 vertices, must have responsibility 25.  Since, the
second hull is non-concentric with the first, by the same argument used 
in Theorem~\ref{thm:singer}, we can always remove one of the vertices 
from the second hull such that the outer two hulls are non-concentric.
This implies that we can create an optimal drawing of $K_9$ whose
outer two hulls are non-concentric, a contradiction of
Theorem~\ref{thm:optimal_k9}.  

Hence, the outer two hulls must be triangular. 
\end{proof}

\begin{theorem}\label{thm:inner}
If $\reccr{K_{10}} = 61$ then an optimal drawing of $K_{10}$ will
consist of two nested triangles containing a convex quadrilateral.
\end{theorem}
\begin{proof}
\figbox[r]{\mbox{ \begin{tabular}{|l|l|}\hline
Crossing & Count \\ \hline
\Xng{gw}{bb}    & 3 \\
\Xng{gw}{gb}    & 3 \\
\Xng{gw}{rb}    & 3 \\ 
\hline
\Xng{rw}{gg}    & 3 \\
\Xng{rw}{bb}    & 3 \\
\Xng{rw}{rg}    & 3 \\ 
\Xng{rw}{rb}    & 3 \\
\hline
\Xng{rw}{gb}    & 6 \\
\Xng{rb}{bw}    &   \\ 
\Xng{gb}{bw}    &   \\ 
\Xng{rg}{gg}    &   \\ \hline
Total    & 27 \\ \hline
\end{tabular}}}{table}{tab:inner}{}
By Lemma~\ref{lem:tri_hull} the outer two hulls of the optimal 
drawing $K_{10}$ must be triangles.  We must still account for the
four internal vertices.  If the four vertices form a convex quadrilateral
then we are done; otherwise, assume the tenth vertex is inside the third
nested triangle.

Colour the tenth vertex white.  Now count the number of red-white and
green-white edge crossings, starting with the green-white edge
crossings.  Each green-white edge must cross into the blue triangle;
multiplying by three yields a total of three \Xng{gw}{bb}\ crossings.
By the $K_5$ principle there are three \Xng{gw}{gb}\ crossings.  Each
blue vertex has three incident red-blue edges that partition the green
\figbox[l]{\boxeps{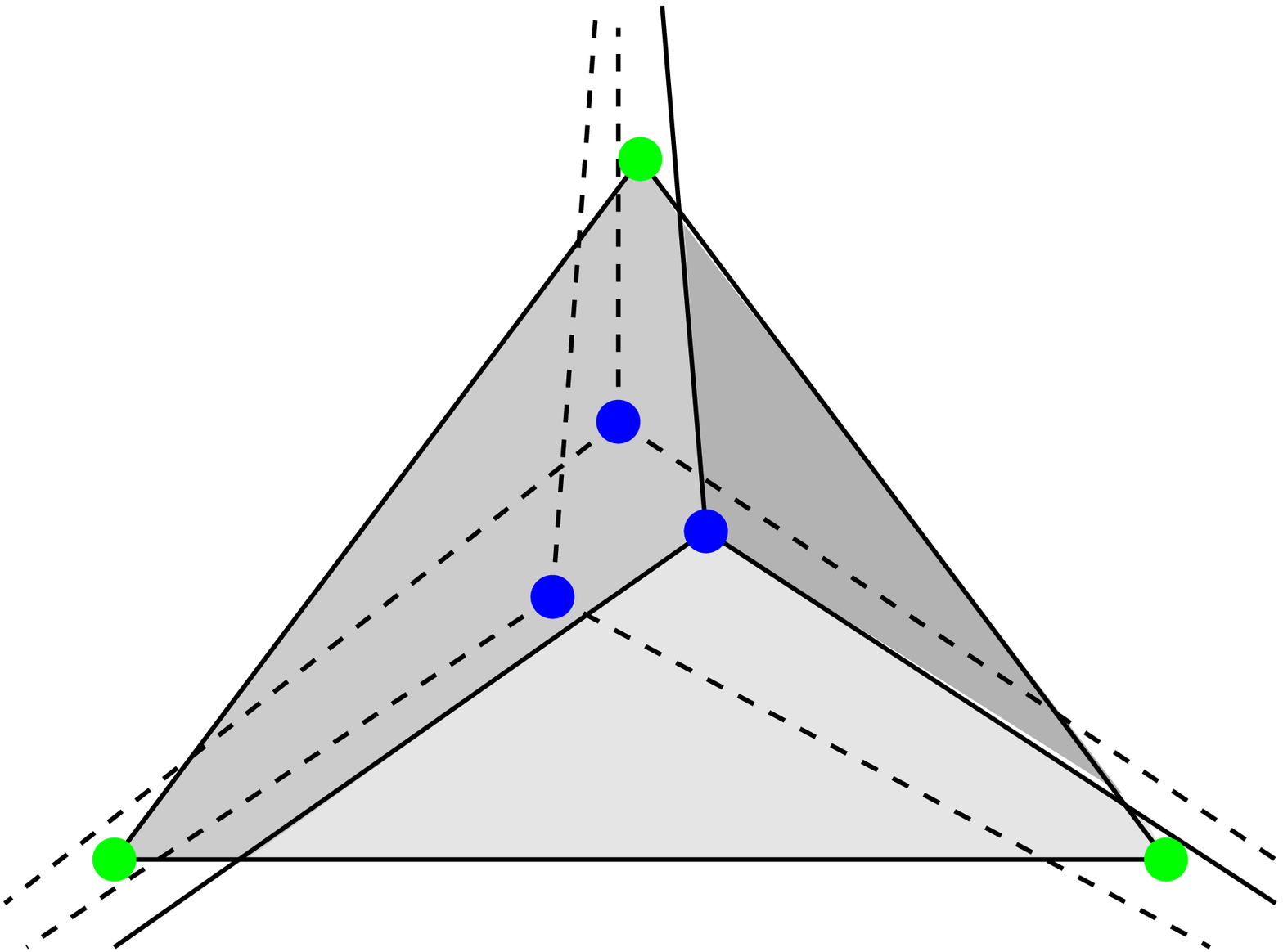}{0.25}}{figure}{fig:tripart}{}
triangle into three regions.  The white vertex must be in one of the
regions; by the Barrier argument there is at least one \Xng{gw}{rb}\ 
crossing per blue vertex.   The total of the green-white edge crossings
sums to nine.

Each red-white edge must cross into both the green and blue triangles,
totaling six edge crossings.  By the $K_5$ principle,
there are three \Xng{rw}{rg}\ crossings and three rw-rb crossings.
This gives an additional 12 crossings.

By Lemma~\ref{lem:rw-bw-x} there are at least six additional crossings
of the \Xng{rw}{gb}, \Xng{rb}{bw}, \Xng{gb}{bw}\, and \Xng{rg}{gg} type, 
of which at least three are \Xng{rw}{gb}\ crossings.

Altogether, the number of white and \Xng{rg}{gg} crossings is 27.
Since $\reccr{K_9} = 36$, the number of edge crossings in the drawing
of $K_{10}$ with the white vertex in the blue triangle is, $36 + 27 =
63 > 61$.
\end{proof}

\begin{theorem}\label{thm:near_main_result}
$\reccr{K_{10}} > 61$.
\end{theorem}
\begin{proof}
By way of contradiction assume that $\reccr{K_{10}} = 61$.
By Theorem~\ref{thm:inner} the inner hull
must be a convex quadrilateral.  Repeat the argument from
Theorem~\ref{thm:inner} disregarding the \Xng{rw}{bb}\ and 
\Xng{gw}{bb}\ edge crossings (because there is no blue triangle).  
This gives us an initial count of $63 - 6 = 57$ edge crossings.  Let
the entire inner convex quadrilateral be coloured blue.  Inside the
\figbox[l]{\boxeps{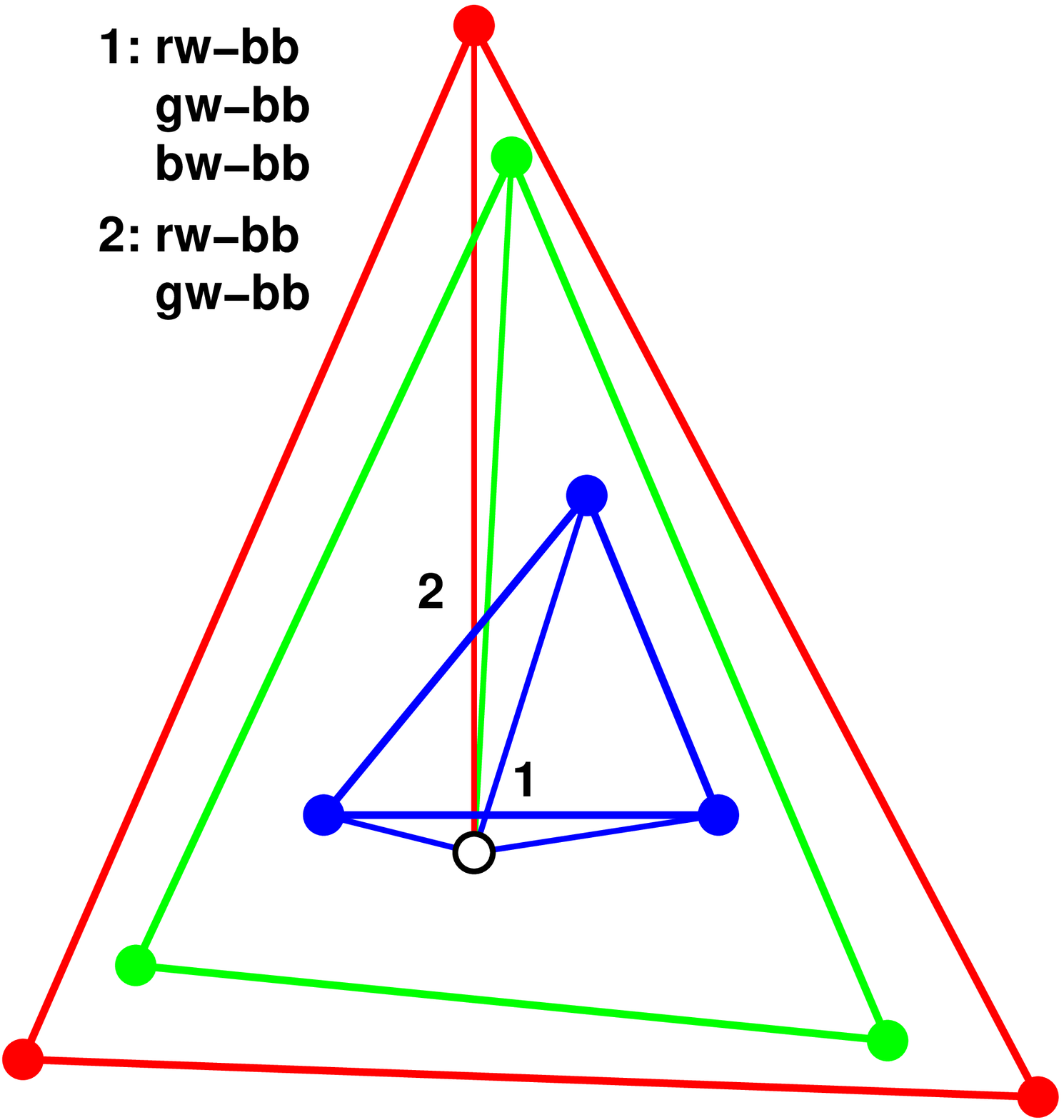}{0.25}}{figure}{fig:phantom}{}
quadrilateral there will be one \Xng{bb}{bb}\ crossing (the
diagonals).  Furthermore, since the quadrilateral is neither concentric
with the red triangle nor the green triangle, there will be a minimum
of two \Xng{rb}{bb}\ edge crossings and two \Xng{gb}{bb}\ edge
crossings.  Summing the edge crossings yields $57 + 5 = 62 > 61$.
\end{proof}

\begin{theorem}\label{thm:main_result}
$\reccr{K_{10}} = 62$.
\end{theorem}
\begin{proof}
Singer's rectilinear drawing of $K_{10}$ with 62 edge crossings
\cite{Si71} is exhibited in \cite[p. 142]{Ga86}, and hence
$\reccr{K_{10}} \leq 62$.  By Theorem~\ref{thm:near_main_result} 
$\reccr{K_{10}} \geq 62$.  The result follows.
\end{proof}

An even stronger statement can be made.  Just as in the case of $K_9$,
the outer two hulls of an optimal rectilinear drawing of $K_{10}$ must
be triangles.
These properties could be useful, just as in the case
of $K_{10}$, for determing the rectilinear crossing number of $K_{11}$.

Theorem~\ref{thm:main_result} enables us to improve the lower bound in 
equations~(\ref{equ:rectasymptotics}) and~(\ref{equ:oldasymptotics}).

\section{Asymptotic Lower Bounds}\label{sec:stephsection}
Given $\reccr{K_a}$ for a fixed $a$, one can derive lower bounds for
all $\reccr{K_n}$, $n>a$.  Any complete subgraph of $a$ vertices drawn
from a rectilinear drawing of $K_n$ will include at least
$\reccr{K_a}$ crossings. There are $\binom{n}{a}$ complete subgraphs
of size $a$.  Each crossing consists of four vertices and each will be
included in all other subgraphs containing the same four vertices. The
number of such subgraphs that share four given vertices is 
$\binom{n-4}{a-4}$.  Guy~\cite{Gu60}, Richter and Thomassen~\cite{RiTh97},
and Scheinerman and Wilf~\cite{ScWi94} each use this argument to show
that

\begin{equation}
\reccr{K_n} \geq \reccr{K_a} 
\binom{n}{a} / \binom{n-4}{a-4} \ .
\end{equation}

Scheinerman and Wilf~\cite{ScWi94} show that this can be rearranged to
get

\begin{equation}
\frac{\reccr{K_n}}{\binom{n}{4}} \geq \frac{\reccr{K_a}}{\binom{a}{4}} \ .
\end{equation}

Thus, one obtains a general lower bound for $\reccr{K_n}$ from any
known $\reccr{K_a}$.  Since $\reccr{K_{10}} = 62$ and $\binom{10}{4} =
210$, one gets

\begin{equation}
\forall n \geq 10, \ 
\frac{\reccr{K_n}}{\binom{n}{4}} \geq \frac{62}{210} \approx 0.2952 \ .
\end{equation}

This raises the lower bound for $\reccr{K_{11}}$ to 98.  We conjecture
$\reccr{K_{11}} = 102$.  Since crossing numbers are integers, each
lower bound can be slightly increased by taking its ceiling.  Thus,

\begin{equation}
\reccr{K_n} \geq \left\lceil \reccr{K_a}
\binom{n}{a} / \binom{n-4}{a-4} \right\rceil \ .
\label{equ:stephequation}
\end{equation}

If one sets $a=n-1$, equation~(\ref{equ:stephequation}) gives a
recursive definition whose recursive ceilings provide an improved lower
bound for $\reccr{K_n}$.  For example, one finds that, $\reccr{K_{400}}
\geq 315356975$. This leads to a general lower bound of

\begin{equation}
\lim_{n\rightarrow\infty}\frac{\reccr{K_n}}{\binom{n}{4}} \geq 
      \frac{315356975}{\binom{400}{4}} = \frac{315356975}{1050739900} 
      \approx 0.3001 \ .
\end{equation}

As $n$ increases, the limit converges.  Whenever $\reccr{K_{a^\prime}}$
is discovered for a new $a^\prime$, one can find an improved lower
bound for a general $\reccr{K_n}$, $n>a^\prime$.

Consequently, $\reccr{K_n}$ can be bound from below by using the
technique describe here and from above by the drawing described by
Brodsky, Durocher, and Gethner in~\cite{BrDuGe00} to achieve the
following lower and upper bounds:

{\small
\begin{equation}
0.3001 \approx \frac{315356975}{1050739900} \leq
\lim_{n\rightarrow\infty}\frac{\reccr{K_n}}{\binom{n}{4}} \leq 
\frac{6467}{16848} \approx 0.3838 \ .
\end{equation}}

\section{Conclusion}\label{futurework}
\subsection{Current and Future Work}
The flavour of finding the crossing number of a graph, particularly in
a rectilinear drawing, is similar to that of determining properties
of line arrangements in the plane; this area is well known to be
delicate and difficult. Therefore, one expects improvements to occur at
a slow rate and specific instances of the problem for small graphs to
be hard, though interesting.

An approach that has proved quite useful is to catalogue all
inequivalent drawings of a given graph. With such a catalogue one can
determine many specific properties of small graphs; see, for example,
\cite{GrHa90,HaTh96}. In particular, to find the crossing number or
rectilinear crossing number of $K_n$ and $K_{m,n}$, one can adopt a
brute-force computational approach to find exact values of the crossing
number for small graphs.  Such an approach is currently underway for
determining $\reccr{K_n}$ by Applegate, Cook, Dash, and
Dean~\cite{De00}, where not only will they independently confirm that
$\reccr{K_{10}}=62$, but they will determine exact values of
$\reccr{K_n}$ for other values of $n\geq 11$ as well.

In fact, when each new value of $\reccr{K_n}$ is found, the lower bounds
in equation (\ref{equ:oldasymptotics}) and equation
(\ref{equ:stephequation}) will improve by way of the technique given in
Section \ref{sec:stephsection}. For example, we have seen that
$\reccr{K_{11}}\geq 98$.  There exists a rectilinear drawing of
$K_{11}$ with 102 edge crossings \cite{Je71,ScWi94}; by \cite{ArRi88},
$\reccr{K_{11}}$ is even.  Therefore, $\reccr{K_{11}}\in\{98,100,102\}$.
If $\reccr{K_{11}}=100$ or 102 then the lower bound in equation
(\ref{equ:stephequation}) becomes .30544 or .31085 respectively.
Similarly, the best drawing of $K_{12}$ known to date has 156 edge
crossings \cite{Je71}; if $\reccr{K_{12}}=156$, then the lower bound
reaches .31839.

Clearly, finding exact values for $\reccr{K_n}$ for any value of
$n$ will make relatively large improvements on the asymptotic lower
bounds for the determination of $\overline{\nu}^*$.

\subsection{Open Problems}
We mention a small subset of open problems that arose from our
investigations.
\begin{enumerate}
\item We know from \cite{Gu72} that if $\regcr{K_n} = \reccr{K_n}$ then
      the convex hull of any optimal rectilinear drawing of $K_n$ is a 
      triangle.  Prove that the convex hull of any optimal rectilinear 
      drawing of $K_n$ is a triangle.

\item Given a rectilinear drawing of $G$, the \myem{planar
      subdivision of $G$} is the graph obtained by adding vertices (and
      corresponding adjacencies) at each of the edge crossings of the
      particular drawing of $G$. Is the planar subdivision of any
      rectilinear drawing of $K_n$ necessarily 3-connected?   This
      question was also posed by Nate Dean.

\item Does there exist an optimal rectilinear drawing of $K_n$, for some 
      $n$, such that it does not contain a sub-drawing that is an optimal
      rectilinear drawing of $K_{n-1}$?  Furthermore, does there exist
      some $n$ for which none of the optimal rectilinear drawings of
      $K_n$ contain a sub-drawing that is an optimal rectilinear drawing
      of $K_{n-1}$?

\item Often an optimal rectilinear drawing of $K_n$ is not unique.
      For a given $n$, how many optimal drawings of $K_n$ are there?
      For what values of $n$ is the optimal drawing unique?

\item Let $G$ be an arbitrary graph.  What is the complexity of 
      determining $\reccr{G}$? Similarly, where in the complexity
      hierarchy does the determination of $\reccr{K_n}$ live? Recall
      that for a not-necessarily-rectilinear drawing of $G$, the
      general problem is known to be NP-complete \cite{GaJo83}. For
      some thoughts on such problems, see \cite{Bi91}.

\item We have seen that $\reccr{K_{11}}\in\{98,100,102\}$ and believe
      $\reccr{K_{11}}$ to be 102. Give a combinatorial proof.

\item Finally, in the spirit of the present paper we feel compelled to
      mention the following problem, for which we sincerely apologize.
      Since 1970 it has been known that $\regcr{K_{7,7}}\in\{77,79,81\}$
      \cite{Kl70}.  What is the final answer?
\end{enumerate}

\section{Acknowledgments}
Crossing number problems are easy to state but notoriously and
profoundly difficult to solve. Throughout our investigations we
discovered the inadequacy of simply searching the existing literature
databases and subsequently reading papers. In particular, much of what
is known or claimed to be known can only be discovered by communicating
directly with those who are acquainted with the crossing number realm.
For this reason, we thank Mike Albertson, Nate Dean, Richard Guy, Heiko
Harborth, Joan Hutchinson, David Kirkpatrick, Ji{\v{r}}{\'\i} 
Matou{\v{s}}ek, Nick Pippenger, David Singer, and Herb Wilf, all of whom 
shared their insights with us.

\bibliography{biblio}

\appendix
\section{Other $K_9$ Drawings}\label{sec:k9_app}
As before, colour the outer hull red, the second hull green, and the 
vertices inside the second hull blue.

\begin{lemma}\label{lem:guilty}
If the first hull of a rectilinear drawing of $K_9$ is a triangle,
and the second hull has six vertices, then the drawing has more than
36 crossings.
\end{lemma}
\begin{proof}
This drawing is coloured by only two colours: red and green.  By the
$K_5$ principle there are $\binom{6}{2} = 15$ \Xng{rg}{rg}\ crossings.
Since the six green vertices comprise the second hull, there are
$\binom{6}{4} = 15$ \Xng{gg}{gg}\ crossings.  The 30 crossings counted
so far include all except the \Xng{rg}{gg}\ crossings.

\figbox[r]{\boxeps{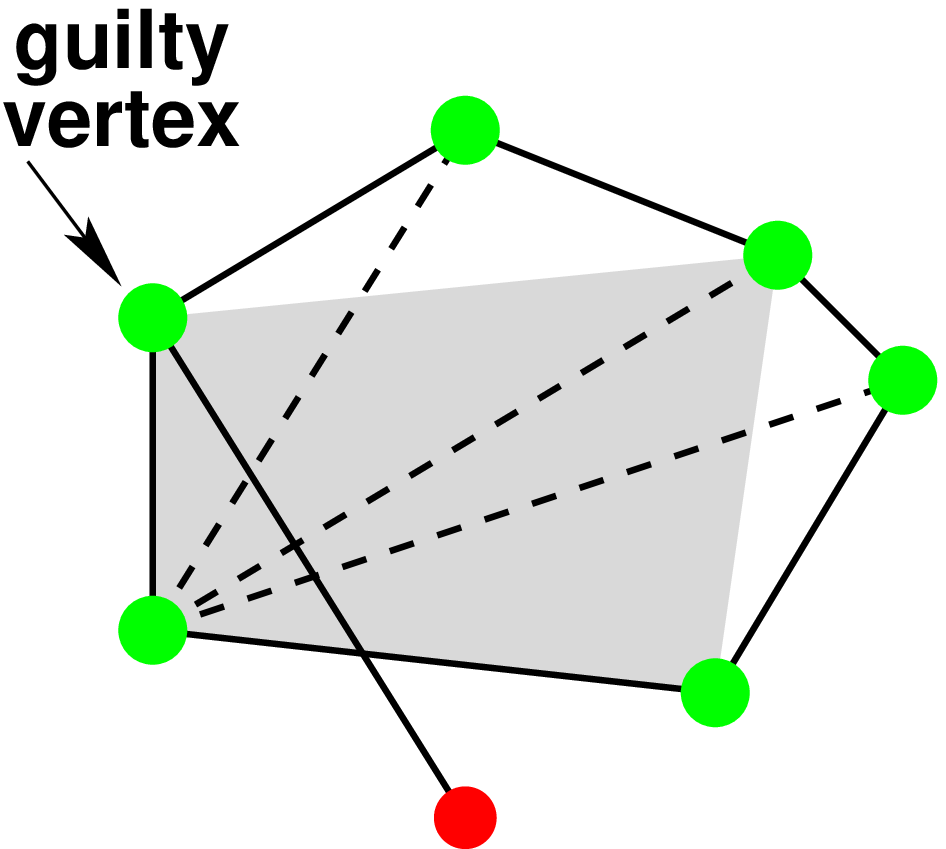}{0.25}}{figure}{fig:guilty}{}
We now consider the \Xng{rg}{gg}\ crossings.  Select four of the green
vertices; these form a convex quadrilateral and at least one green vertex,
the guilty vertex, has a red-green edge that intersects the quadrilateral.
This edge partitions the green hull into two parts with one green vertex
on one side of the green hull and three on the other, or two on each side.
In the former case the red-green edge crosses three green-green edges that
are incident on the single vertex.  In the latter case, the red-green
edge intersects four green-green edges that are incident on the two
green vertices in one of the partitions.  In both cases, there is an
additional \Xng{rg}{gg}\ crossing due to the red-green edge crossing
an edge of the quadrilateral.  Hence, at minimum four \Xng{rg}{gg}\
crossings are due to the single red-green edge.  Since there are at
least three guilty vertices in a hull on six vertices.  There must be
at least 12 \Xng{rg}{gg}\ crossings.

Therefore, the total number of crossings is at least $42 > 36$.
\end{proof}

\begin{lemma}\label{lem:guilty2}
If the first hull of a rectilinear drawing of $K_9$ is a triangle,
and the second convex hull has five vertices, then the drawing has
more than 36 crossings.
\end{lemma}
\begin{proof}
As before, the single vertex inside the second hull is coloured blue.
By the $K_5$ principle there are $\binom{5}{2} = 10$ \Xng{rg}{rg}\ 
crossings and $\binom{5}{1} = 5$ \Xng{rg}{rb}\ crossings.  By the same
argument used in the previous lemma there are $\binom{5}{4} = 5$
\Xng{gg}{gg}\ crossings.  There are at least five \Xng{gb}{gg}\ 
crossings.  Thus, we reach a count of 25 crossings without having
considered the \Xng{rb}{gg}, \Xng{rg}{gg}, and \Xng{rg}{gb}\ 
crossings.

We count the \Xng{rg}{gg}, and \Xng{rg}{gb}\ crossings by the guilty
vertex argument used in the previous lemma.  A hull on five vertices
will have at least two guilty vertices.  Each guilty vertex is
responsible for at least three \Xng{rg}{gg}\ crossings and, by the
Barrier argument, at least one \Xng{rg}{gb}\ crossing.  This yields an
additional eight crossings, bringing the total up to 33.

\figbox[l]{\mbox{ \begin{tabular}{|l|l|}\hline
Crossing        & Min \\ \hline
\Xng{rg}{rg}    & 10 \\
\Xng{rg}{rb}    & 5  \\
\Xng{gg}{gg}    & 5  \\
\Xng{gb}{gg}    & 5  \\ \hline
\Xng{rg}{gg}    & 6  \\
\Xng{rg}{gb}    & 2  \\
\Xng{rb}{gg}    & 5  \\ \hline
Total           & 38 \\ \hline
\end{tabular}}}{table}{tab:app_k9_v5}{Crossings}
Finally, consider the \Xng{rb}{gg}\ crossings.  At least three occur
from the red-blue edges having to cross into the green hull.  By the
containment argument, at least one of these three edges has to cross two
of the green-green diagonals within the green hull.  This brings up the
total to at least five \Xng{rb}{gg}\ crossings.  Adding this to the 
running total yields $38 > 36$.
\end{proof}

\end{document}